\documentclass[preprintnumbers,twocolumn,prx,aps,showpacs,10pt,superscriptaddress,floatfix,longbibliography]{revtex4-1}
\usepackage{enumerate}
\usepackage{graphicx}
\usepackage{amsmath,amssymb,amsthm,enumitem}

\usepackage{color}
\definecolor{LinkColor}{RGB}{199,21,133}
\usepackage[colorlinks=true,linkcolor=blue,anchorcolor=blue,citecolor=blue,urlcolor=blue]{hyperref}
\usepackage[polutonikogreek,english]{babel}

\usepackage{hyperref}

\usepackage{epstopdf}
\usepackage{type1cm}
\usepackage{multirow}
\usepackage{times}

\newcommand{\headline}[1]{\textit{\textcolor{magenta}{#1.}}}

\def\scro{\mathcal{O}}

\begin{document}

\title{Quantum extraordinary-log universality of boundary critical behavior}
\author{Yanan Sun}
\author{Jian-Ping Lv}
\email{jplv2014@ahnu.edu.cn}
\affiliation{Department of Physics and Anhui Key Laboratory of Optoelectric Materials Science and Technology, Key Laboratory of Functional Molecular Solids, Ministry of Education, Anhui Normal University, Wuhu, Anhui 241000, China}
\date{\today}
\begin{abstract}
The recent discovery of extraordinary-log universality has generated intense interest in classical and
quantum boundary critical phenomena. Despite tremendous efforts, the existence of quantum extraordinary-log universality remains extremely controversial. Here, by utilizing quantum Monte Carlo simulations, we study the quantum edge criticality of a two-dimensional Bose-Hubbard model featuring emergent bulk criticality. On top of an insulating bulk, the open edges experience a Kosterlitz-Thouless-like transition into the superfluid phase when the hopping strength is sufficiently enhanced on edges. At the bulk critical point, the open edges exhibit the special, ordinary, and extraordinary critical phases. In the extraordinary phase, logarithms are involved in the finite-size scaling of two-point correlation and superfluid stiffness, which admit a classical-quantum correspondence for the extraordinary-log universality. Thanks to modern quantum emulators for interacting bosons in lattices, the edge critical phases might be realized in experiments.
\end{abstract}


\date{\today}

\maketitle

\section{Introduction}
Scaling and universality are pillars of modern critical phenomena~\cite{stanley1999scaling}. In the paradigm of criticality, the two-point correlation $g(r)$ decays as the power law~\cite{domb1996critical,stanley1999scaling,sachdev2007quantum,fernandez2013random}
\begin{equation}\label{tp2}
g(r) \sim r^{2-(d+z)-\eta}
\end{equation}
with the spatial distance $r$, where $d$, $z$ and $\eta$ are respectively spatial dimension, dynamic critical exponent and anomalous dimension.

Boundary critical behavior (BCB) refers to the critical phenomena occurring on boundaries of a critical bulk~\cite{binder1974surface,Ohno1984,landau1989monte,diehl1997theory,pleimling2004critical,deng2005surface,deng2006bulk,Dubail2009,zhang2017unconventional,ding2018engineering,Weber2018,weber2019nonordinary} and relates to a rich variety of state-of-the-art concepts~\cite{cardy2004boundary,grover2012quantum,parker2018topological,poland2019conformal,Liu2021,dantchev2022critical,Andrei_2020}. Recently, in the context of BCB, the extraordinary-log universality (ELU) was predicted by Metlitski for the classical three-dimensional (3D) O($N$) model with $2 \leq N < N_c$, where $N_c$ is an upper bound~\cite{metlitski2020boundary}. For ELU the boundary two-point correlation $g(r)$ decays logarithmically with $r$ as~\cite{metlitski2020boundary}
\begin{equation}\label{tp1}
g(r) \sim [{\rm ln}(r)]^{-\hat{\eta}},
\end{equation}
where $\hat{\eta}$ is only dependent on $N$. Shortly afterwards, much attention was devoted to the BCB in classical~\cite{toldin2020boundary,Hu2021,padayasi2021extraordinary,ToldinMetlitski2021extraordinary,ToldinSurface2021,Zhang2022Surface,Zou2022Surface} and quantum~\cite{Weber2021,ding2021special,Zhu2021Exotic,Yu2021Conformal,Xu2021Persistent} systems.

Evidence for \textit{classical} ELU was obtained from the Monte Carlo simulations of Heisenberg and XY models~\cite{toldin2020boundary,Hu2021,ToldinMetlitski2021extraordinary}. Inspired by the studies using magnetic fluctuations at different Fourier modes to explore precise finite-size scaling (FSS)~\cite{Wittmann2014,Flores-Sola2016} as well as the two-length scenarios for high-dimensional Ising models~\cite{papathanakos2006finite,grimm2017geometric,zhou2018random,FangComplete,lv2021finite,Fang2021} and deconfined criticality~\cite{shao2016quantum}, an alternative scaling formula of $g(r)$ was conjectured for ELU~\cite{Hu2021}. This conjecture was based on the fact that the critical magnetic fluctuations at zero and smallest non-zero modes scale as $L^2 [{\rm ln}(L)]^{-\hat{q}}$ and $L^2 [{\rm ln}(L)]^{-\hat{\eta}}$, with the critical exponents $\hat{q}$ and $\hat{\eta}=\hat{q}+1$, respectively. This observation can be related to the FSS of $g(r)$ as~\cite{Hu2021}
\begin{equation}\label{tp3}
g(r) \sim \begin{cases} [{\rm ln}(r)]^{-\hat{\eta}}, & {\rm ln}(r) \le  \scro [({\rm ln}(L))^{\hat{q}/\hat{\eta}}],  \\
[{\rm ln}(L)]^{-\hat{q}},  & {\rm ln}(r) \ge \scro [({\rm ln}(L))^{\hat{q}/\hat{\eta}}].
\end{cases}
\end{equation}
With the concept ``unwrapping''~\cite{heydenreich2017progress,grimm2017geometric,bet2021detecting}, a geometric explanation of two-length scenario was introduced based on unwrapped correlation length~\cite{grimm2017geometric,Fang2021,deng2022unwrapped}. The two exponents $\hat{q}$ and $\hat{\eta}$ were also observed in the classical ELU at an emergent O(2) critical point~\cite{Zhang2022Surface}. Eq.~(\ref{tp3}) {\it formally} agrees with (\ref{tp1}) on the FSS of $g(r)$ in the $r \rightarrow \infty$ limit.

Quantum edge criticality (QEC) has been extensively studied in the two-dimensional dimerized antiferromagnetic quantum (2D-DAQ) Heisenberg and XXZ models, which are prototype models for O(3) and O(2) criticality~\cite{zhang2017unconventional,ding2018engineering,Weber2018,weber2019nonordinary,Weber2021,ding2018engineering,ding2021special,Zhu2021Exotic}, respectively. On one hand, the dangling edges of 2D-DAQ spin-1/2 and spin-1 Heisenberg models harbor the non-ordinary criticality~\cite{ding2018engineering,Weber2018,weber2019nonordinary,Weber2021}, where the critical exponents in magnetic sector are almost compatible with O(3) special transition~\cite{ding2018engineering,Weber2018}. The numerical results for scaling dimension $\Delta_n$ ($\Delta_v$) of N\'{e}el (valence bond solid) order were compared~\cite{Weber2021} to the field-theoretic prediction~\cite{jian2021continuous}
\begin{equation}\label{tp0}
\Delta_n-1/2=\epsilon_n \,\,\, \mbox{and} \,\,\, \Delta_v-1/2=-3\epsilon_n
\end{equation}
with $\Delta_\phi-3/2 = -\epsilon_n$, where $\Delta_\phi \approx 1.187$~\cite{deng2005surface} is the scaling dimension of spin order in O(3) ordinary universality. For the spin-1/2 case, the results do not agree with Eq.~(\ref{tp0}) but conform with the scaling relation $3\Delta_n+\Delta_v=2$. For the spin-1 case, the estimate $\Delta_v \approx -2$ is roughly compatible with the theory of extraordinary-power phase~\cite{metlitski2020boundary}, hence in sharp contrast to Eq.~(\ref{tp0}) and the theory of ELU. On the other hand, the non-dangling edges of 2D-DAQ spin-1/2 Heisenberg model host the ordinary phase, special transition and {\it long-range ordered} extraordinary phase
~\cite{ding2018engineering,Weber2018,ding2021special}. Moreover, the 2D-DAQ spin-1 XXZ model may exhibit the extraordinary-log criticality, yet this observation does not hold for the spin-1/2 case~\cite{Zhu2021Exotic}.

\begin{figure}
	\includegraphics[height=5.5cm,width=8cm]{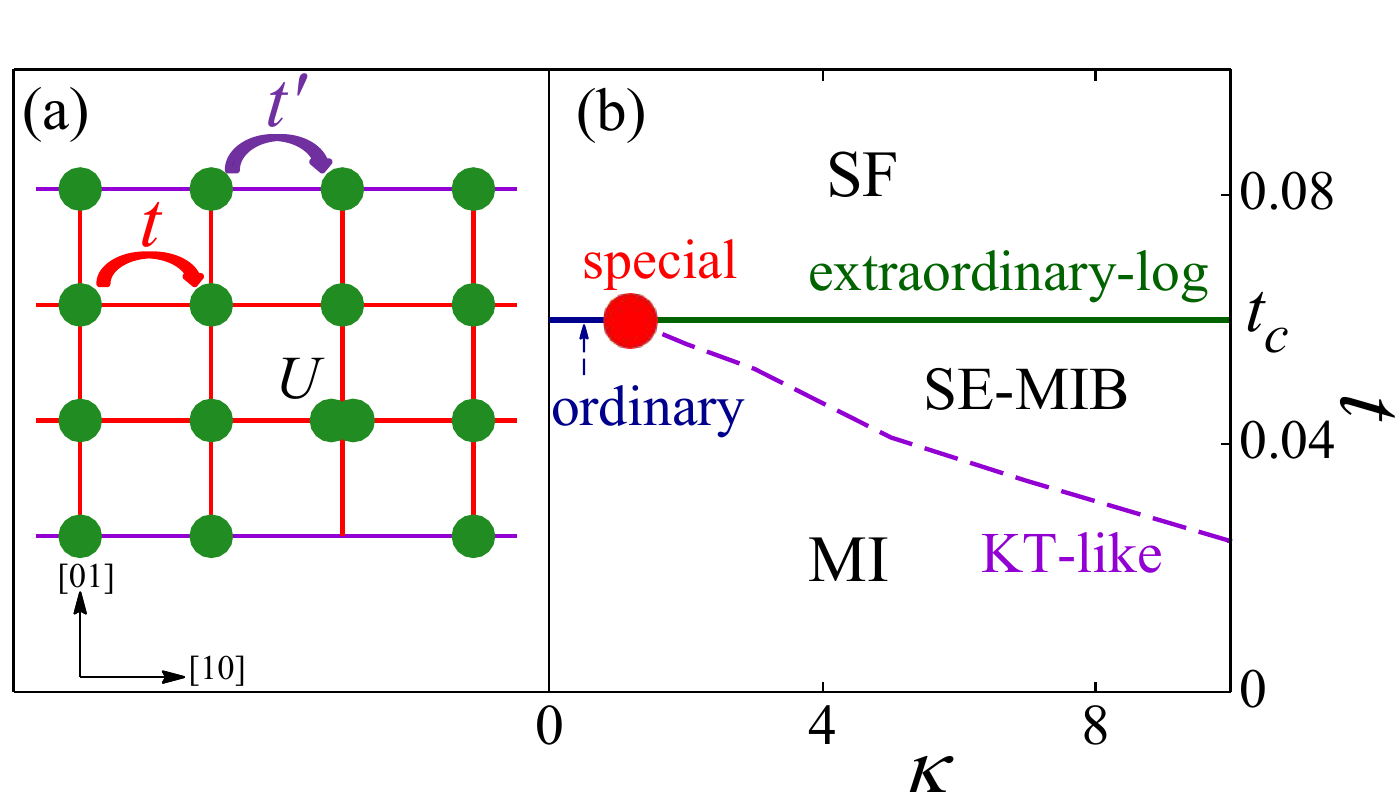}
	\caption{Model and ground-state phase diagram. (a) Definition of open-edge Bose-Hubbard model, where $t$ and $t'$ are hopping amplitudes and $U$ denotes onsite repulsion. (b) The phase diagram in terms of $t$ and the edge hopping enhancement $\kappa$, including a phase with superfluid edges and Mott insulating bulk (SE-MIB) as well as the phases of bulk-edge superfluid (SF) and Mott insulator (MI). These phases are separated by the Kosterlitz-Thouless-like (KT-like), extraordinary-log and ordinary critical lines that are terminated at the multi-critical special transition point.}~\label{fig1}
\end{figure}

Hence, despite the tremendous efforts devoted to the BCB of quantum antiferromagnets, the existence of quantum ELU remains extremely controversial. Moreover, as indicated in Ref.~\cite{metlitski2020boundary}, the existing results can not form a self-contained picture for the classical-quantum correspondence of BCB and failed to realize quantum ELU. Here, we switch to interacting bosons and show that the open-edge Bose-Hubbard model hosts quantum ELU. This conclusion is based on the logarithmic FSS of two-point correlation and superfluid stiffness for extraordinary phase as well as an overall classical-quantum correspondence for various critical phases. The sharp difference from the BCB of XXZ antiferromagnet~\cite{Zhu2021Exotic} reflects the sensitivity of BCB to geometric settings and local operators.

In the following, we focus on the open-edge Bose-Hubbard model and explore the quantum O(2) BCB of the model. Section~\ref{msec1} defines the open-edge Bose-Hubbard model and presents its ground-state phase diagram. Section~\ref{msec2} introduces the methodology adopted throughout present study. Section~\ref{msec3} presents Monte Carlo data and scaling analyses. A summary is finally given in Sec.~\ref{msec4}.

\section{Model and ground-state phase diagram}\label{msec1}
We consider the square-lattice Bose-Hubbard model at unit boson filling with the Hamiltonian
\begin{equation}
\hat{H}= - \sum\limits_{\langle ij \rangle} t_{ij} (\hat{b}_{i}^\dagger \hat{b}_{j} + \hat{b}_{j}^\dagger \hat{b}_{i}) + \frac{U}{2} \sum\limits_{i}  \hat{n}_i(\hat{n}_i-1),
\label{Ham1}
\end{equation}
where $\hat{b}_{i}^\dagger$ and $\hat{b}_{i}$ are respectively bosonic creation and annihilation operators at site $i$, and $\hat{n}_i=\hat{b}_{i}^\dagger \hat{b}_{i}$. $t_{ij}$ denotes the amplitude of the nearest-neighbor hopping between $i$ and $j$, and $U>0$ represents onsite repulsion. The first summation runs over pairs of nearest neighboring sites while the second summation is over sites. We set $U=1$ as energy unit.

As illustrated by Fig.~\ref{fig1}(a), we define our model for BCB by setting open and periodic boundary conditions along [01] and [10] directions, respectively. Hence, a pair of open edges are specified. The hopping amplitude $t_{ij}=t'$ on open edges is distinguished from $t_{ij}=t$ in bulk. The edge hopping enhancement is parameterized by $\kappa=(t'-t)/t$.

At $\kappa=0$, model~(\ref{Ham1}) reduces to the standard Bose-Hubbard model at unit boson filling~\cite{Fisher1989}, which has an {\it emergent} O(2) quantum critical point separating the Mott insulating and superfluid phases. This critical point features Lorentz invariance with $z=1$. The present authors and coworkers have given an estimate for the quantum critical point as $t_c=0.059\,729\,1(8)$~\cite{Xu2019}, which agrees with the literature result $t_c=0.059\,74(3)$~\cite{capogrosso2008monte}.

We explore quantum phases of model (\ref{Ham1}) by FSS and the results are summarized as a ground-state phase diagram in Fig.~\ref{fig1}(b). There is a phase, dubbed SE-MIB, that features superfluid edges on top of Mott insulating bulk. Moreover, there are three critical edge phases at $t_c$: the ordinary, special and extraordinary-log phases. Scaling behaviors of edge critical phases are described in Table~\ref{tablesum}~\footnote{Logarithmic corrections may emerge for the KT-like transition and the SE-MIB phase~\cite{kosterlitz2016kosterlitz}}.

\begin{table}
\caption{Leading scaling behaviors of the edge two-point correlation $g(L/2)$ and the superfluid stiffness $\rho_s$
in critical phases.}
\centering
\begin{tabular}{c|c|c}
\hline
Critical phase      \,     & \, $g(L/2)$ \,   & \, $\rho_s$ \, \\
\hline
special  & $L^{-\eta}$, \, $\eta \approx 0.65$ &  $L^{-1}$      \\
KT-like   & $L^{-\eta}$, \, $\eta = 1/4$ & $L^{-1}$     \\
SE-MIB   & $L^{-\eta}$, \, $\eta$ $\in$ (0, 1/4)  & $L^{-1}$     \\
ordinary    & $L^{-\eta}$, \, $\eta \approx 2.438$      & $L^{-1}$    \\
extraordinary   &$ \, \, [{\rm ln} (L)]^{-\hat{q}}$, \, $\hat{q} \approx 0.59$ \, & \, $L^{-1} {\rm ln} (L)$   \\
\hline
\end{tabular}
\label{tablesum}
\end{table}

\section{Methodology}\label{msec2}
We apply the Prokof'ev-Svistunov-Tupitsyn worm quantum Monte Carlo algorithm~\cite{prokofev1998exact,prokofev1998worm} to simulate model (\ref{Ham1}) in the imaginary-time path integral representation. The maximum side length of the square lattice is up to $L=192$.
The inverse temperature is set as $\beta=L$, which is in line with $z=1$. We study the special, ordinary and extraordinary phases at $t_c=0.059\,729\,1$ by varying $\kappa$, and explore the KT-like transition for $t<t_c$. In particular, we analyze the extraordinary phase in a broad parameter regime.

Analyses of the FSS involving ${\rm ln}(L)$ may be ``notoriously difficult''~\cite{Grassberger2003}.
We perform the analyses using least-squares fits. Following standard criterion, we prefer the fits with $\chi^2/{\rm DF} \sim 1$, where $\chi^2$ is the Chi squared and DF denotes the degree of freedom. We also examine the stability against varying $L_{\rm min}$, which represents the minimum side length involved in fitting.

\section{Results}\label{msec3}
\subsection{Special transition}
\begin{figure}
\includegraphics[height=9cm,width=9cm]{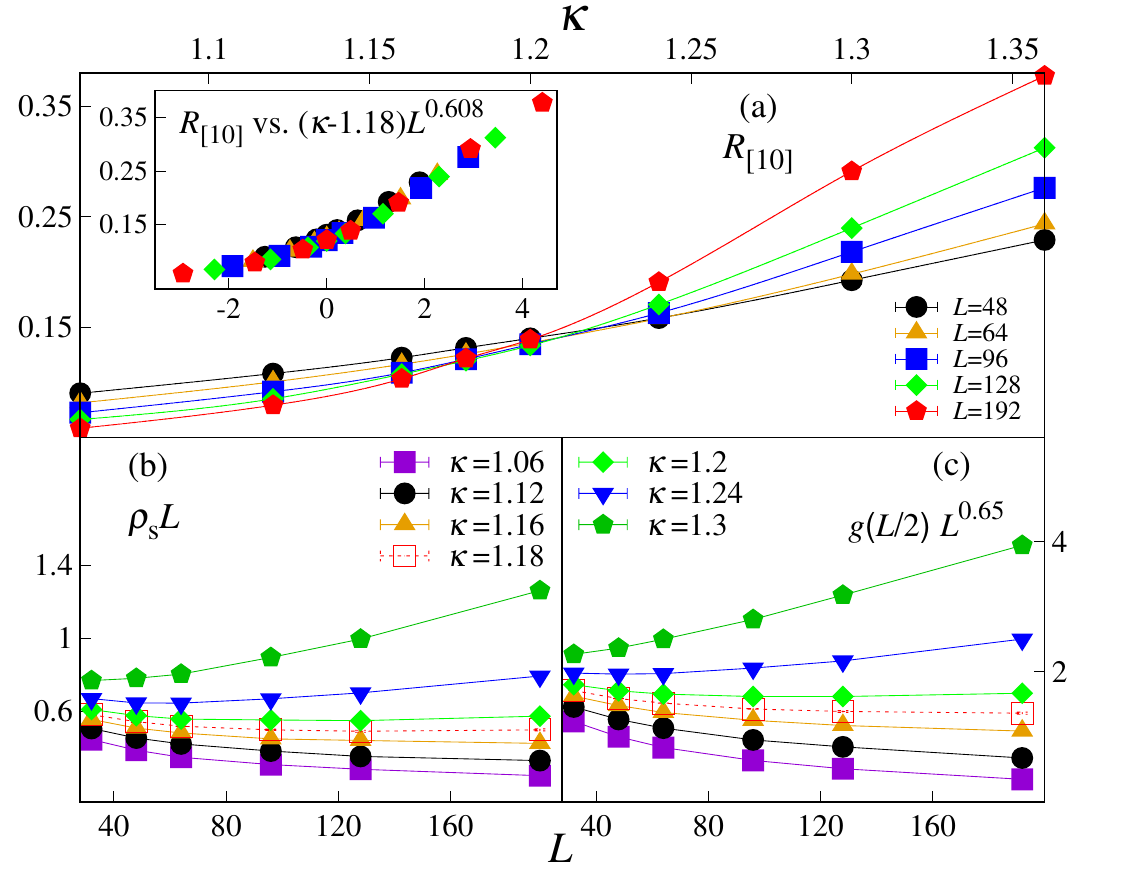}
\caption{Special transition. (a) Winding probability $R_{[10]}$ versus $\kappa$. The inset displays $R_{[10]}$ versus $(\kappa-\kappa_c)L^{y_t}$ with $\kappa_c=1.18$ and $y_t=0.608$. (b) Scaled superfluid stiffness $\rho_sL$ versus $L$. (c) Scaled two-point correlation $g(L/2)L^{4-2y_h}$ with $y_h=1.675$.}~\label{fig2}
\end{figure}

We detect the special transition by tuning $\kappa$ at $t=t_c$. We sample the
winding probability $R_{[10]}=\langle \mathcal{R}_{[10]} \rangle$, where $\mathcal{R}_{[10]}=1$ if
there exists at least a particle line winding around the periodic [10] direction of square lattice.
The winding probability is dimensionless and obeys the FSS $R_{[10]}=\widetilde{R}_{[10]} (\epsilon L^{y_t})$, where $\epsilon=\kappa-\kappa_c$ represents the deviation from the critical point $\kappa_c$, and $y_t$ relates to the correlation length exponent $\nu$ by $y_t=1/\nu$. $R_{[10]}$ is useful for locating critical points~\cite{Xu2019}. Expanding $\widetilde{R}_{[10]}$ and incorporating corrections to scaling, we obtain
\begin{equation}
	R_{[10]}= R^c_{[10]}+\sum\limits_j a_j \epsilon^j L^{j y_t}+\sum\limits_m b_m L^{-\omega_m},
	\label{fitRx}
\end{equation}
where $R^c_{[10]}$ is somewhat universal, $a_j$ $(j=1,2,\dots)$ and $b_m$ $(m=1,2,\dots)$ are non-universal,
and $\omega_m$ represents exponents for corrections. We show $R_{[10]}$ versus $\kappa$ in Fig.~\ref{fig2}(a), where a scaling invariance point is nearly at $\kappa \approx 1.2$. We fit $R_{[10]}$ data with $L=48, 64, 96, 128$ and $192$ to Eq.~(\ref{fitRx}).
We observe $\omega_1 \approx 1.4$, which is larger than $\omega_1 \approx 0.789$ of 3D O(2) value~\cite{guida1998critical} and $\omega_1=1$ from boundary irrelevant fields~\cite{toldin2020boundary}. The correction with $\omega_1 \le 1$ is either absent or weak. Hence, we also perform fits without correction term and monitor the effects of corrections by examining the stability of fits upon gradually increasing $L_{\rm min}$. We obtain $\kappa_c=1.206(7)$ and $y_t=0.44(8)$ with $\chi^2/{\rm DF} \approx 4.6$ for $L_{\rm min}=64$, $\kappa_c=1.184(6)$ and $y_t=0.4(1)$ with $\chi^2/{\rm DF}\approx 0.9$ for $L_{\rm min}=96$, as well as $\kappa_c=1.175(5)$ and $y_t=0.8(3)$ with $\chi^2/{\rm DF} \approx 0.2$ for $L_{\rm min}=128$. Next, by fixing $y_t$ at the estimate $y_t=0.608$ for the special transition of classical O(2) model~\cite{deng2005surface}, we obtain $\kappa_c=1.197(2)$, $1.180(3)$ and $1.175(7)$ with $\chi^2/{\rm DF} \approx 4.6$, $1.1$ and $0.3$, for $L_{\rm min}=64$, $96$ and $128$, respectively. When $y_t=0.58$ is fixed, we obtain close estimates, which are detailed in Appendix C. By comparing all these fits, we finally estimate $\kappa_c=1.18(2)$. For illustrating the single-variable function $\widetilde{R}_{[10]}$ together with the estimates of $\kappa_c$ and $y_t$, we plot $R_{[10]}$ versus $\epsilon L^{y_t}$ in Fig.~\ref{fig2}(a) with $\kappa_c=1.18$ and $y_t=0.608$, where finite-size corrections are already negligible for large systems.

Further evidence comes from the FSS of the superfluid stiffness $\rho_s$, which is defined as~\cite{Pollock1987} $\rho_s= \langle \mathcal{W}_{[10]}^2 \rangle/(2t'\beta)$ through the fluctuations of the winding number $\mathcal{W}_{[10]}$ along the [10] direction of square lattice. At $\kappa_c$, $\rho_s$ should scale as $\rho_s \sim L^{2-(d+z)}$. This scaling behavior is verified by Fig.~\ref{fig2}(b) with $d=2$ and $z=1$: as $L \rightarrow \infty$, $\rho_s L$ is asymptotically a constant for $\kappa \leq \kappa_c$, but bends upwards for $\kappa> \kappa_c$.

We consider the two-point correlation $g(L/2)$ at the largest distance $r_{[10]}=L/2$ along an open edge, which is estimated from the random walks of the two defects in worm quantum Monte Carlo simulations. More descriptions and benchmarks for this estimator are presented in Appendix B. Figure~\ref{fig2}(c) shows that the result at $\kappa_c$ is compatible with the critical scaling behavior $g(L/2) \sim L^{-0.65}$, yet deviates when $\kappa \ne \kappa_c$. The scaling behavior at $\kappa_c$ is accounted for by the O(2) special universality with the exponent $y_h \approx 1.675$~\cite{deng2005surface,Zhang2022Surface,Zou2022Surface,unpublished}, as $g(L/2) \sim L^{2y_h-4}$.

\begin{figure}
	\includegraphics[height=9cm,width=9cm]{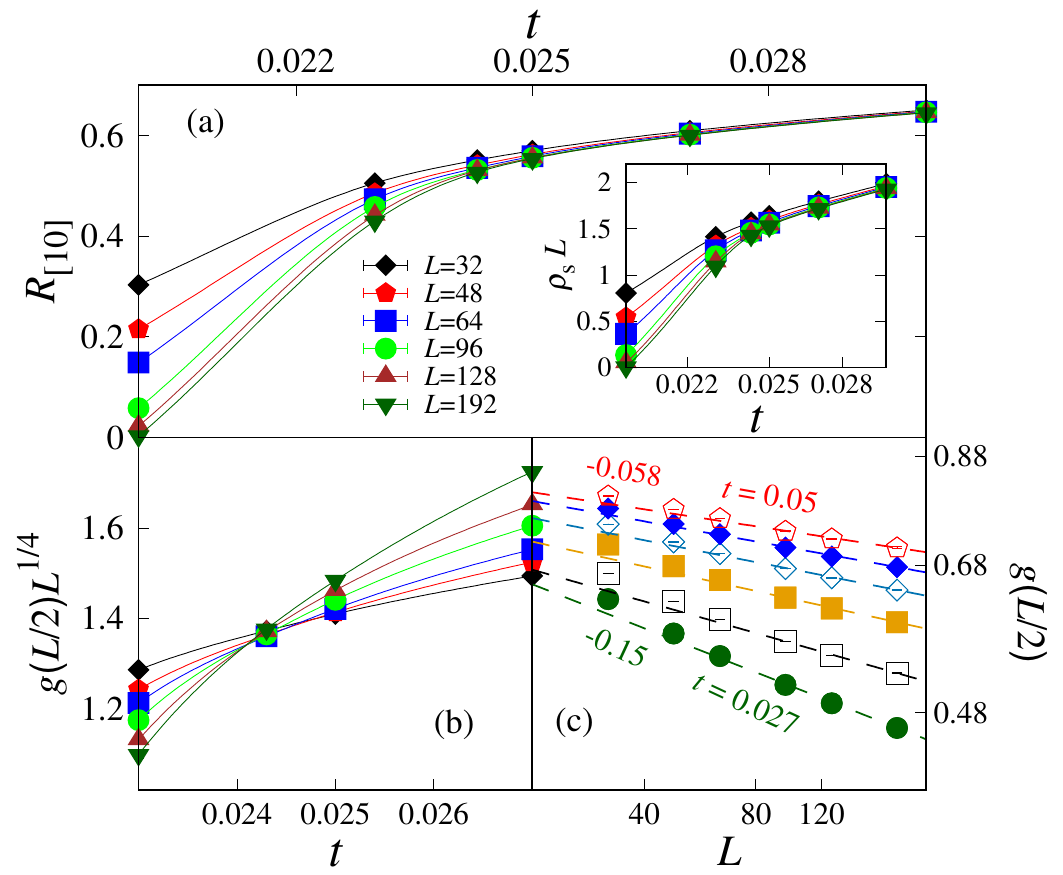}
	\caption{KT-like criticality ($\kappa=10$). (a) Winding probability $R_{[10]}$ versus $t$. The inset displays the scaled superfluid stiffness $\rho_s L$. (b) Scaled two-point correlation $g(L/2)L^{1/4}$ versus $t$. (c) Log-log plot of $g(L/2)$ versus $L$.}~\label{fig3}
\end{figure}

\subsection{KT-like criticality}
Figure~\ref{fig3}(a) shows $R_{[10]}$ versus $t$ for $\kappa=10$. Around $t_x \approx 0.023$, $R_{[10]}$ varies drastically. For $t > t_x$, $R_{[10]}$ extrapolates to a nontrivial value in the $L \rightarrow \infty$ limit, which is dependent on $t$. Meanwhile, the superfluid stiffness scales as $\rho_s \sim  L^{-1}$. These observations indicate a regime of critical phase.

The KT-like criticality is evidenced by the anomalous dimension $\eta$. Figure~\ref{fig3}(b) demonstrates that, at $t_{\rm KT} \approx t_x$, $g(L/2)$ scales as $g(L/2) \sim L^{2-(d+z)-\eta}$ with $d=1$, $z=1$ and $\eta=1/4$. The value $1/4$ is consistent with that of the KT transition in 2D XY model~\cite{kosterlitz1974critical}. For $t>t_{\rm KT}$, we fit $g(L/2)$ to the formula $g(L/2)\sim L^{-\eta}$ of leading scaling. The fits are illustrated by Fig.~\ref{fig3}(c) and detailed in Appendix C. In particular, for $t=0.027$ and $0.05$, we obtain $\eta = 0.150(2)$ and $0.058(4)$ respectively, with $\chi^2/\rm{DF} \approx 1.0$ and $L_{\rm min}=96$. The continuously varying exponent $\eta$ is reminiscent of the low-temperature critical phase of 2D XY model~\cite{kosterlitz2016kosterlitz}.

\subsection{Ordinary critical phase}
\begin{figure}
	\includegraphics[height=5.5cm,width=9cm]{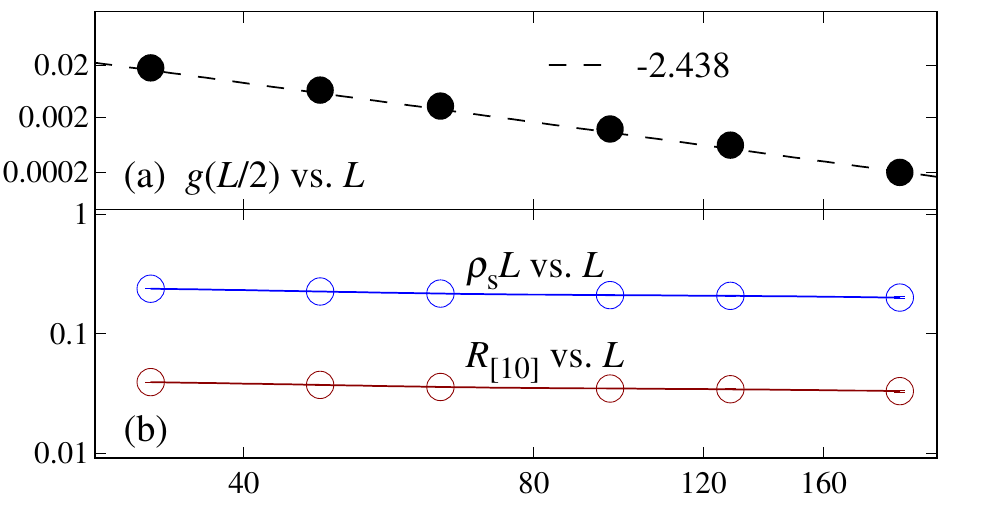}
	\caption{Ordinary critical phase ($\kappa=0.4$). (a) Log-log plot of two-point correlation $g(L/2)$ versus $L$. The slope $-2.438$ relates to $2y_h-4$ with $y_h=0.781$. (b) Log-log plot of scaled superfluid stiffness $\rho_s L$ and winding probability $R_{[10]}$ versus $L$.}~\label{fig4}
\end{figure}

Corresponding to classical O(2) BCB, the small-$\kappa$ side of special transition may fall into the ordinary critical universality class. For $\kappa=0.4$, Fig.~\ref{fig4} demonstrates that $g(L/2)$ scales as $L^{2-(d+z)-\eta}$ with $\eta \approx 2.438$, $d=1$ and $z=1$. The value of $\eta$ relates to $y_h=0.781(2)$~\cite{deng2005surface} of the O(2) BCB by $\eta=4-2y_h$. As $L \rightarrow \infty$, $\rho_s L$ and $R_{[10]}$ tend to be independent of $L$. These scaling behaviors indicate the existence of the O(2) quantum ordinary universality.

\subsection{Extraordinary-log critical phase}
\begin{figure}
	\includegraphics[height=9cm,width=9cm]{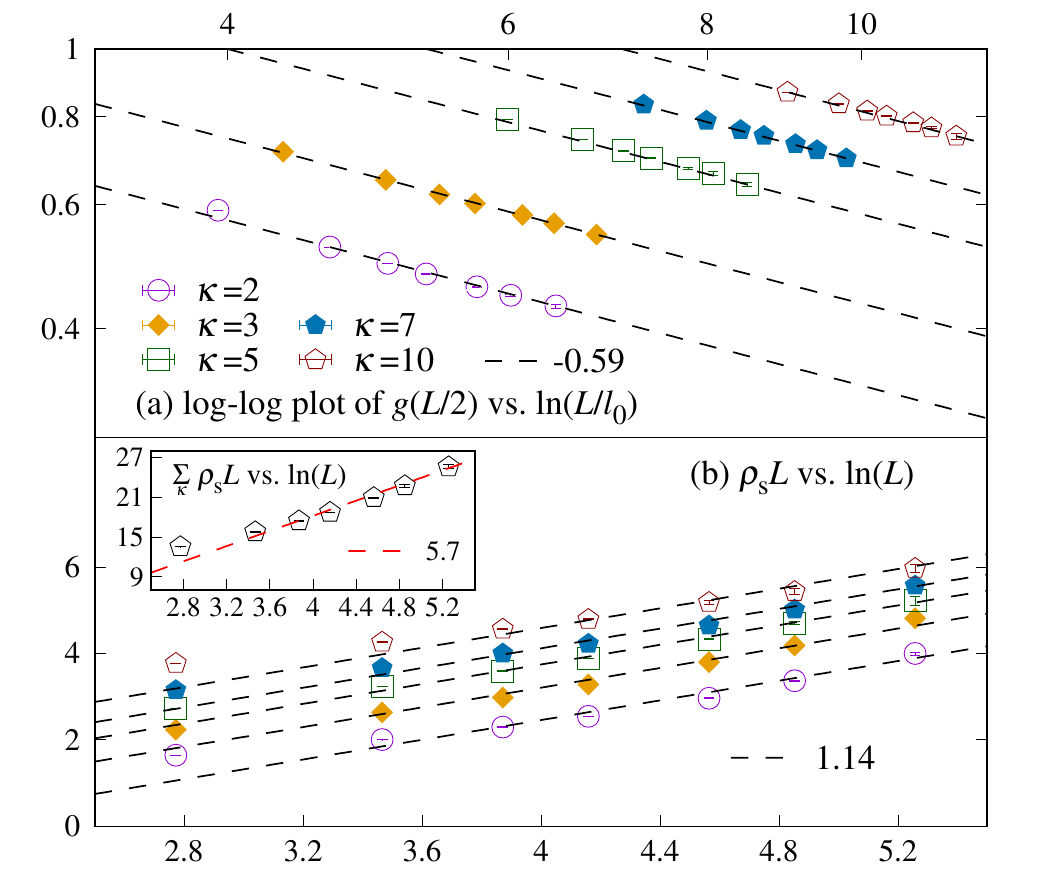}
	\caption{Extraordinary-log critical phase. (a) Log-log plot of two-point correlation $g(L/2)$ versus ${\rm ln}(L/l_0)$, where the values of $l_0$ come from preferred fits. The slope $-0.59$ relates to $-\hat{q}$. (b) Scaled superfluid stiffness $\rho_s L$ versus ${\rm ln}(L)$. Inset: the summation of $\rho_s L$ over $\kappa$. The slopes $1.14$ and $5.7$ denote $b$ in Eq.~(\ref{fitrhosE}) and $5b$, respectively.}~\label{fig5}
\end{figure}

To explore the extraordinary phase, we make use of a broad parameter regime in the large-$\kappa$ side of special transition. In the ELU, $g(L/2)$ scales as~\cite{metlitski2020boundary}
\begin{equation}
	g(L/2) =a [{\rm ln}(L/l_0)]^{-\hat{q}},
	\label{fitg2LE}
\end{equation}
where $l_0$ is a reference length and $a$ denotes a non-universal constant. For the classical XY model, this scaling form was verified and $\hat{q}=0.59(2)$ was estimated~\cite{Hu2021}. Close values of $\hat{q}$ were obtained for the classical ELU of O(2) model~\cite{ToldinMetlitski2021extraordinary} and emergent O(2) criticality~\cite{Zhang2022Surface,Zou2022Surface}. We perform fits for $g(L/2)$ according to Eq.~(\ref{fitg2LE}) and obtain $0.3 \lessapprox \hat{q} \lessapprox 0.7$ for $\kappa=2$, $3$, $5$ and $7$. We observe that $l_0$ decreases significantly as $\kappa$ increases. These features conform to the observations for classical ELU in Ref.~\cite{Hu2021}. When $\hat{q}=0.59$ is fixed, we achieve, for each $\kappa$, stable fitting results for $l_0$ and $a$. Instance results of $l_0$ include $l_0=0.31(3)$, $0.21(1)$, $0.04(4)$, $0.0108(5)$ and $0.002(1)$ with $\chi^2 /{\rm DF} \approx 0.3$, $1.8$, $0.9$, $0.7$ and $0.5$, for $\kappa=2$, $3$, $5$, $7$ and $10$, respectively. The power-law dependence of $g(L/2)$ on ${\rm ln}(L/l_0)$ is illustrated by Fig.~\ref{fig5}(a).

From Fig.~\ref{fig5}(b), we find that $\rho_s L$ roughly obeys the logarithmic scaling formula
\begin{equation}
	\rho_s L =b {\rm ln}(L)+c
	\label{fitrhosE}
\end{equation}
with universal $b \approx 1.1$ and non-universal $c$. Preferred fits are achieved in deep extraordinary regime. With $L_{\rm max}=192$, we obtain $b=1.14(3)$, $1.15(3)$ and $1.1(1)$ with $\chi^2/{\rm DF} \approx 0.9$, $2.8$ and $0.7$, for $\kappa=5$, $7$ and $10$, respectively. We also perform fits to $\sum_{\kappa} \rho_s L =5b {\rm ln}(L)+C$ ($C$ is a fitting parameter) where the summation runs over the set $\{2,3,5,7,10\}$ of $\kappa$. For $L_{\rm min}=64$, we obtain reasonably good results as $5b=5.8(2)$ and $C=-5.3(7)$ with $\chi^2/{\rm DF} \approx 2.0$ and $L_{\rm max}=192$, as well as $5b=5.6(2)$ and $C=-4.6(8)$ with $\chi^2/{\rm DF} \approx 0.8$ and $L_{\rm max}=128$. These fits are consistent and finally yield $5b=5.7(3)$, which relates to $b=1.14(6)$. By contrast, the logarithmic divergence of $\rho_s L$ is absent in the paradigm of criticality, as illustrated for special transition [Fig.~\ref{fig2}(b)] and ordinary critical phase [Fig.~\ref{fig4}(b)], and does not emerge in the KT-like criticality [Fig.~\ref{fig3}(a)]. The logarithmic FSS (\ref{fitrhosE}) with unit exponent and universal coefficient resembles that of the helicity modulus in classical XY and Heisenberg models~\cite{metlitski2020boundary,toldin2020boundary,Hu2021}.

\section{Summary}\label{msec4}
The extensive ongoing activities in the search for quantum ELU are restricted to dimerized antiferromagnets, for which conclusive evidence remains unavailable. Here, we switch to interacting bosons by formulating an open-edge Bose-Hubbard model and demonstrate the emergence of quantum ELU. An edge superfluid phase is observed on top of an insulating bulk. When the bulk is at the emergent quantum critical point, the special, ordinary and extraordinary-log critical phases emerge on open edges. In the extraordinary-log critical phase, the leading FSS for the largest-distance two-point correlation and scaled superfluid stiffness are logarithmic. By an overall classical-quantum correspondence of O($2$) BCB as well as the universal behavior of logarithmic FSS in the extraordinary phase, we provide complementary evidence for the existence of quantum ELU. As the Bose-Hubbard model can be accessed by quantum emulators with ultracold bosons in optical lattices~\cite{jaksch1998cold,greiner2002quantum,baier2016extended,yang2020cooling}, our results indicate a possible experimental scheme for realizing ELU.

\begin{acknowledgments}
One of us (J.P.L.) wishes to warmly thank Youjin Deng for the collaboration in earlier related studies. The present work has been supported by the National Natural Science Foundation of China (under Grant Nos. 12275002, 11975024, and 11774002) and the Education Department of Anhui.
\end{acknowledgments}

\bibliography{papers}

\begin{thebibliography}{67}%
\makeatletter
\providecommand \@ifxundefined [1]{%
 \@ifx{#1\undefined}
}%
\providecommand \@ifnum [1]{%
 \ifnum #1\expandafter \@firstoftwo
 \else \expandafter \@secondoftwo
 \fi
}%
\providecommand \@ifx [1]{%
 \ifx #1\expandafter \@firstoftwo
 \else \expandafter \@secondoftwo
 \fi
}%
\providecommand \natexlab [1]{#1}%
\providecommand \enquote  [1]{``#1''}%
\providecommand \bibnamefont  [1]{#1}%
\providecommand \bibfnamefont [1]{#1}%
\providecommand \citenamefont [1]{#1}%
\providecommand \href@noop [0]{\@secondoftwo}%
\providecommand \href [0]{\begingroup \@sanitize@url \@href}%
\providecommand \@href[1]{\@@startlink{#1}\@@href}%
\providecommand \@@href[1]{\endgroup#1\@@endlink}%
\providecommand \@sanitize@url [0]{\catcode `\\12\catcode `\$12\catcode
  `\&12\catcode `\#12\catcode `\^12\catcode `\_12\catcode `\%12\relax}%
\providecommand \@@startlink[1]{}%
\providecommand \@@endlink[0]{}%
\providecommand \url  [0]{\begingroup\@sanitize@url \@url }%
\providecommand \@url [1]{\endgroup\@href {#1}{\urlprefix }}%
\providecommand \urlprefix  [0]{URL }%
\providecommand \Eprint [0]{\href }%
\providecommand \doibase [0]{http://dx.doi.org/}%
\providecommand \selectlanguage [0]{\@gobble}%
\providecommand \bibinfo  [0]{\@secondoftwo}%
\providecommand \bibfield  [0]{\@secondoftwo}%
\providecommand \translation [1]{[#1]}%
\providecommand \BibitemOpen [0]{}%
\providecommand \bibitemStop [0]{}%
\providecommand \bibitemNoStop [0]{.\EOS\space}%
\providecommand \EOS [0]{\spacefactor3000\relax}%
\providecommand \BibitemShut  [1]{\csname bibitem#1\endcsname}%
\let\auto@bib@innerbib\@empty
\bibitem [{\citenamefont {Stanley}(1999)}]{stanley1999scaling}%
  \BibitemOpen
  \bibfield  {author} {\bibinfo {author} {\bibfnamefont {H.~E.}\ \bibnamefont
  {Stanley}},\ }\bibfield  {title} {\enquote {\bibinfo {title} {Scaling,
  universality, and renormalization: Three pillars of modern critical
  phenomena},}\ }\href
  {https://journals.aps.org/rmp/abstract/10.1103/RevModPhys.71.S358} {\bibfield
   {journal} {\bibinfo  {journal} {Rev. Mod. Phys.}\ }\textbf {\bibinfo
  {volume} {71}},\ \bibinfo {pages} {S358} (\bibinfo {year}
  {1999})}\BibitemShut {NoStop}%
\bibitem [{\citenamefont {Domb}(1996)}]{domb1996critical}%
  \BibitemOpen
  \bibfield  {author} {\bibinfo {author} {\bibfnamefont {C.}~\bibnamefont
  {Domb}},\ }\href
  {https://www.taylorfrancis.com/books/mono/10.1201/9781482295269/critical-point-domb}
  {\emph {\bibinfo {title} {The critical point: a historical introduction to
  the modern theory of critical phenomena}}}\ (\bibinfo  {publisher} {CRC
  Press},\ \bibinfo {year} {1996})\BibitemShut {NoStop}%
\bibitem [{\citenamefont {Sachdev}(2007)}]{sachdev2007quantum}%
  \BibitemOpen
  \bibfield  {author} {\bibinfo {author} {\bibfnamefont {S.}~\bibnamefont
  {Sachdev}},\ }\href
  {https://onlinelibrary.wiley.com/doi/abs/10.1002/9780470022184.hmm108} {\emph
  {\bibinfo {title} {Quantum phase transitions}}}\ (\bibinfo  {publisher}
  {Wiley Online Library},\ \bibinfo {year} {2007})\BibitemShut {NoStop}%
\bibitem [{\citenamefont {Fern{\'a}ndez}\ \emph {et~al.}(2013)\citenamefont
  {Fern{\'a}ndez}, \citenamefont {Fr{\"o}hlich},\ and\ \citenamefont
  {Sokal}}]{fernandez2013random}%
  \BibitemOpen
  \bibfield  {author} {\bibinfo {author} {\bibfnamefont {R.}~\bibnamefont
  {Fern{\'a}ndez}}, \bibinfo {author} {\bibfnamefont {J.}~\bibnamefont
  {Fr{\"o}hlich}}, \ and\ \bibinfo {author} {\bibfnamefont {A.~D.}\
  \bibnamefont {Sokal}},\ }\href
  {https://link.springer.com/book/10.1007%2F978-3-662-02866-7} {\emph {\bibinfo
  {title} {Random walks, critical phenomena, and triviality in quantum field
  theory}}}\ (\bibinfo  {publisher} {Springer, Berlin},\ \bibinfo {year}
  {2013})\BibitemShut {NoStop}%
\bibitem [{\citenamefont {Binder}\ and\ \citenamefont
  {Hohenberg}(1974)}]{binder1974surface}%
  \BibitemOpen
  \bibfield  {author} {\bibinfo {author} {\bibfnamefont {K}~\bibnamefont
  {Binder}}\ and\ \bibinfo {author} {\bibfnamefont {P.~C.}\ \bibnamefont
  {Hohenberg}},\ }\bibfield  {title} {\enquote {\bibinfo {title} {Surface
  effects on magnetic phase transitions},}\ }\href
  {https://journals.aps.org/prb/abstract/10.1103/PhysRevB.9.2194} {\bibfield
  {journal} {\bibinfo  {journal} {Phys. Rev. B}\ }\textbf {\bibinfo {volume}
  {9}},\ \bibinfo {pages} {2194} (\bibinfo {year} {1974})}\BibitemShut
  {NoStop}%
\bibitem [{\citenamefont {Ohno}\ and\ \citenamefont {Okabe}(1984)}]{Ohno1984}%
  \BibitemOpen
  \bibfield  {author} {\bibinfo {author} {\bibfnamefont {K.}~\bibnamefont
  {Ohno}}\ and\ \bibinfo {author} {\bibfnamefont {Y.}~\bibnamefont {Okabe}},\
  }\bibfield  {title} {\enquote {\bibinfo {title} {The 1/n expansion for the
  extraordinary transition of semi-infinite system},}\ }\href {\doibase
  10.1143/PTP.72.736} {\bibfield  {journal} {\bibinfo  {journal} {Prog. Theor.
  Phys.}\ }\textbf {\bibinfo {volume} {72}},\ \bibinfo {pages} {736--745}
  (\bibinfo {year} {1984})}\BibitemShut {NoStop}%
\bibitem [{\citenamefont {Landau}\ \emph {et~al.}(1989)\citenamefont {Landau},
  \citenamefont {Pandey},\ and\ \citenamefont {Binder}}]{landau1989monte}%
  \BibitemOpen
  \bibfield  {author} {\bibinfo {author} {\bibfnamefont {D.~P.}\ \bibnamefont
  {Landau}}, \bibinfo {author} {\bibfnamefont {R.}~\bibnamefont {Pandey}}, \
  and\ \bibinfo {author} {\bibfnamefont {K.}~\bibnamefont {Binder}},\
  }\bibfield  {title} {\enquote {\bibinfo {title} {Monte carlo study of surface
  critical behavior in the $xy$ model},}\ }\href
  {https://journals.aps.org/prb/abstract/10.1103/PhysRevB.39.12302} {\bibfield
  {journal} {\bibinfo  {journal} {Phys. Rev. B}\ }\textbf {\bibinfo {volume}
  {39}},\ \bibinfo {pages} {12302} (\bibinfo {year} {1989})}\BibitemShut
  {NoStop}%
\bibitem [{\citenamefont {Diehl}(1997)}]{diehl1997theory}%
  \BibitemOpen
  \bibfield  {author} {\bibinfo {author} {\bibfnamefont {H.~W.}\ \bibnamefont
  {Diehl}},\ }\bibfield  {title} {\enquote {\bibinfo {title} {The theory of
  boundary critical phenomena},}\ }\href
  {https://www.worldscientific.com/doi/abs/10.1142/S0217979297001751}
  {\bibfield  {journal} {\bibinfo  {journal} {Int. J. Mod. Phys. B}\ }\textbf
  {\bibinfo {volume} {11}},\ \bibinfo {pages} {3503--3523} (\bibinfo {year}
  {1997})},\ \Eprint {http://arxiv.org/abs/cond-mat/9610143}
  {arXiv:cond-mat/9610143 [cond-mat]} \BibitemShut {NoStop}%
\bibitem [{\citenamefont {Pleimling}(2004)}]{pleimling2004critical}%
  \BibitemOpen
  \bibfield  {author} {\bibinfo {author} {\bibfnamefont {M.}~\bibnamefont
  {Pleimling}},\ }\bibfield  {title} {\enquote {\bibinfo {title} {Critical
  phenomena at perfect and non-perfect surfaces},}\ }\href
  {https://iopscience.iop.org/article/10.1088/0305-4470/37/19/R01/meta}
  {\bibfield  {journal} {\bibinfo  {journal} {J. Phys. A: Math. and Gen.}\
  }\textbf {\bibinfo {volume} {37}},\ \bibinfo {pages} {R79} (\bibinfo {year}
  {2004})},\ \Eprint {http://arxiv.org/abs/cond-mat/0402574}
  {arXiv:cond-mat/0402574 [cond-mat]} \BibitemShut {NoStop}%
\bibitem [{\citenamefont {Deng}\ \emph {et~al.}(2005)\citenamefont {Deng},
  \citenamefont {Bl{\"o}te},\ and\ \citenamefont
  {Nightingale}}]{deng2005surface}%
  \BibitemOpen
  \bibfield  {author} {\bibinfo {author} {\bibfnamefont {Y.}~\bibnamefont
  {Deng}}, \bibinfo {author} {\bibfnamefont {H.~W.~J.}\ \bibnamefont
  {Bl{\"o}te}}, \ and\ \bibinfo {author} {\bibfnamefont {M.~P.}\ \bibnamefont
  {Nightingale}},\ }\bibfield  {title} {\enquote {\bibinfo {title} {Surface and
  bulk transitions in three-dimensional $o(n)$ models},}\ }\href
  {https://journals.aps.org/pre/abstract/10.1103/PhysRevE.72.016128} {\bibfield
   {journal} {\bibinfo  {journal} {Phys. Rev. E}\ }\textbf {\bibinfo {volume}
  {72}},\ \bibinfo {pages} {016128} (\bibinfo {year} {2005})},\ \Eprint
  {http://arxiv.org/abs/cond-mat/0504173} {arXiv:cond-mat/0504173 [cond-mat]}
  \BibitemShut {NoStop}%
\bibitem [{\citenamefont {Deng}(2006)}]{deng2006bulk}%
  \BibitemOpen
  \bibfield  {author} {\bibinfo {author} {\bibfnamefont {Y.}~\bibnamefont
  {Deng}},\ }\bibfield  {title} {\enquote {\bibinfo {title} {Bulk and surface
  phase transitions in the three-dimensional o(4) spin model},}\ }\href
  {https://journals.aps.org/pre/abstract/10.1103/PhysRevE.73.056116} {\bibfield
   {journal} {\bibinfo  {journal} {Phys. Rev. E}\ }\textbf {\bibinfo {volume}
  {73}},\ \bibinfo {pages} {056116} (\bibinfo {year} {2006})}\BibitemShut
  {NoStop}%
\bibitem [{\citenamefont {Dubail}\ \emph {et~al.}(2009)\citenamefont {Dubail},
  \citenamefont {Jacobsen},\ and\ \citenamefont {Saleur}}]{Dubail2009}%
  \BibitemOpen
  \bibfield  {author} {\bibinfo {author} {\bibfnamefont {J.}~\bibnamefont
  {Dubail}}, \bibinfo {author} {\bibfnamefont {J.~L.}\ \bibnamefont
  {Jacobsen}}, \ and\ \bibinfo {author} {\bibfnamefont {H.}~\bibnamefont
  {Saleur}},\ }\bibfield  {title} {\enquote {\bibinfo {title} {Exact solution
  of the anisotropic special transition in the o(n) model in two dimensions},}\
  }\href {\doibase 10.1103/PhysRevLett.103.145701} {\bibfield  {journal}
  {\bibinfo  {journal} {Phys. Rev. Lett.}\ }\textbf {\bibinfo {volume} {103}},\
  \bibinfo {pages} {145701} (\bibinfo {year} {2009})},\ \Eprint
  {http://arxiv.org/abs/0909.2949} {arXiv:0909.2949 [cond-mat]} \BibitemShut
  {NoStop}%
\bibitem [{\citenamefont {Zhang}\ and\ \citenamefont
  {Wang}(2017)}]{zhang2017unconventional}%
  \BibitemOpen
  \bibfield  {author} {\bibinfo {author} {\bibfnamefont {L.}~\bibnamefont
  {Zhang}}\ and\ \bibinfo {author} {\bibfnamefont {F.}~\bibnamefont {Wang}},\
  }\bibfield  {title} {\enquote {\bibinfo {title} {Unconventional surface
  critical behavior induced by a quantum phase transition from the
  two-dimensional affleck-kennedy-lieb-tasaki phase to a n{\'e}el-ordered
  phase},}\ }\href
  {https://journals.aps.org/prl/abstract/10.1103/PhysRevLett.118.087201}
  {\bibfield  {journal} {\bibinfo  {journal} {Phys. Rev. Lett.}\ }\textbf
  {\bibinfo {volume} {118}},\ \bibinfo {pages} {087201} (\bibinfo {year}
  {2017})},\ \Eprint {http://arxiv.org/abs/1611.06477} {arXiv:1611.06477
  [cond-mat]} \BibitemShut {NoStop}%
\bibitem [{\citenamefont {Ding}\ \emph {et~al.}(2018)\citenamefont {Ding},
  \citenamefont {Zhang},\ and\ \citenamefont {Guo}}]{ding2018engineering}%
  \BibitemOpen
  \bibfield  {author} {\bibinfo {author} {\bibfnamefont {C.}~\bibnamefont
  {Ding}}, \bibinfo {author} {\bibfnamefont {L.}~\bibnamefont {Zhang}}, \ and\
  \bibinfo {author} {\bibfnamefont {W.}~\bibnamefont {Guo}},\ }\bibfield
  {title} {\enquote {\bibinfo {title} {Engineering surface critical behavior of
  (2+1)-dimensional o(3) quantum critical points},}\ }\href
  {https://journals.aps.org/prl/abstract/10.1103/PhysRevLett.120.235701}
  {\bibfield  {journal} {\bibinfo  {journal} {Phys. Rev. Lett.}\ }\textbf
  {\bibinfo {volume} {120}},\ \bibinfo {pages} {235701} (\bibinfo {year}
  {2018})},\ \Eprint {http://arxiv.org/abs/1801.10035} {arXiv:1801.10035
  [cond-mat]} \BibitemShut {NoStop}%
\bibitem [{\citenamefont {Weber}\ \emph {et~al.}(2018)\citenamefont {Weber},
  \citenamefont {Parisen~Toldin},\ and\ \citenamefont {Wessel}}]{Weber2018}%
  \BibitemOpen
  \bibfield  {author} {\bibinfo {author} {\bibfnamefont {L.}~\bibnamefont
  {Weber}}, \bibinfo {author} {\bibfnamefont {F.}~\bibnamefont
  {Parisen~Toldin}}, \ and\ \bibinfo {author} {\bibfnamefont {S.}~\bibnamefont
  {Wessel}},\ }\bibfield  {title} {\enquote {\bibinfo {title} {Nonordinary edge
  criticality of two-dimensional quantum critical magnets},}\ }\href {\doibase
  10.1103/PhysRevB.98.140403} {\bibfield  {journal} {\bibinfo  {journal} {Phys.
  Rev. B}\ }\textbf {\bibinfo {volume} {98}},\ \bibinfo {pages} {140403(R)}
  (\bibinfo {year} {2018})},\ \Eprint {http://arxiv.org/abs/1804.06820}
  {arXiv:1804.06820 [cond-mat]} \BibitemShut {NoStop}%
\bibitem [{\citenamefont {Weber}\ and\ \citenamefont
  {Wessel}(2019)}]{weber2019nonordinary}%
  \BibitemOpen
  \bibfield  {author} {\bibinfo {author} {\bibfnamefont {L.}~\bibnamefont
  {Weber}}\ and\ \bibinfo {author} {\bibfnamefont {S.}~\bibnamefont {Wessel}},\
  }\bibfield  {title} {\enquote {\bibinfo {title} {Nonordinary criticality at
  the edges of planar spin-1 heisenberg antiferromagnets},}\ }\href
  {https://journals.aps.org/prb/abstract/10.1103/PhysRevB.100.054437}
  {\bibfield  {journal} {\bibinfo  {journal} {Phys. Rev. B}\ }\textbf {\bibinfo
  {volume} {100}},\ \bibinfo {pages} {054437} (\bibinfo {year} {2019})},\
  \Eprint {http://arxiv.org/abs/1906.07051} {arXiv:1906.07051 [cond-mat]}
  \BibitemShut {NoStop}%
\bibitem [{\citenamefont {Cardy}()}]{cardy2004boundary}%
  \BibitemOpen
  \bibfield  {author} {\bibinfo {author} {\bibfnamefont {J.}~\bibnamefont
  {Cardy}},\ }\bibfield  {title} {\enquote {\bibinfo {title} {Boundary
  conformal field theory},}\ }\href@noop {} {\ }\Eprint
  {http://arxiv.org/abs/hep-th/0411189} {arXiv:hep-th/0411189 [cond-mat]}
  \BibitemShut {NoStop}%
\bibitem [{\citenamefont {Grover}\ and\ \citenamefont
  {Vishwanath}()}]{grover2012quantum}%
  \BibitemOpen
  \bibfield  {author} {\bibinfo {author} {\bibfnamefont {T.}~\bibnamefont
  {Grover}}\ and\ \bibinfo {author} {\bibfnamefont {A.}~\bibnamefont
  {Vishwanath}},\ }\bibfield  {title} {\enquote {\bibinfo {title} {Quantum
  criticality in topological insulators and superconductors: Emergence of
  strongly coupled majoranas and supersymmetry},}\ }\href@noop {} {\ }\Eprint
  {http://arxiv.org/abs/1206.1332} {arXiv:1206.1332 [cond-mat]} \BibitemShut
  {NoStop}%
\bibitem [{\citenamefont {Parker}\ \emph {et~al.}(2018)\citenamefont {Parker},
  \citenamefont {Scaffidi},\ and\ \citenamefont
  {Vasseur}}]{parker2018topological}%
  \BibitemOpen
  \bibfield  {author} {\bibinfo {author} {\bibfnamefont {D.~E.}\ \bibnamefont
  {Parker}}, \bibinfo {author} {\bibfnamefont {T.}~\bibnamefont {Scaffidi}}, \
  and\ \bibinfo {author} {\bibfnamefont {R.}~\bibnamefont {Vasseur}},\
  }\bibfield  {title} {\enquote {\bibinfo {title} {Topological luttinger
  liquids from decorated domain walls},}\ }\href
  {https://journals.aps.org/prb/abstract/10.1103/PhysRevB.97.165114} {\bibfield
   {journal} {\bibinfo  {journal} {Phys. Rev. B}\ }\textbf {\bibinfo {volume}
  {97}},\ \bibinfo {pages} {165114} (\bibinfo {year} {2018})},\ \Eprint
  {http://arxiv.org/abs/1711.09106} {arXiv:1711.09106 [cond-mat]} \BibitemShut
  {NoStop}%
\bibitem [{\citenamefont {Poland}\ \emph {et~al.}(2019)\citenamefont {Poland},
  \citenamefont {Rychkov},\ and\ \citenamefont {Vichi}}]{poland2019conformal}%
  \BibitemOpen
  \bibfield  {author} {\bibinfo {author} {\bibfnamefont {D.}~\bibnamefont
  {Poland}}, \bibinfo {author} {\bibfnamefont {S.}~\bibnamefont {Rychkov}}, \
  and\ \bibinfo {author} {\bibfnamefont {A.}~\bibnamefont {Vichi}},\ }\bibfield
   {title} {\enquote {\bibinfo {title} {The conformal bootstrap: Theory,
  numerical techniques, and applications},}\ }\href
  {https://journals.aps.org/rmp/abstract/10.1103/RevModPhys.91.015002}
  {\bibfield  {journal} {\bibinfo  {journal} {Rev. Mod. Phys.}\ }\textbf
  {\bibinfo {volume} {91}},\ \bibinfo {pages} {015002} (\bibinfo {year}
  {2019})},\ \Eprint {http://arxiv.org/abs/1805.04405} {arXiv:1805.04405
  [cond-mat]} \BibitemShut {NoStop}%
\bibitem [{\citenamefont {Liu}\ \emph {et~al.}(2021)\citenamefont {Liu},
  \citenamefont {Shapourian}, \citenamefont {Vishwanath},\ and\ \citenamefont
  {Metlitski}}]{Liu2021}%
  \BibitemOpen
  \bibfield  {author} {\bibinfo {author} {\bibfnamefont {S.}~\bibnamefont
  {Liu}}, \bibinfo {author} {\bibfnamefont {H.}~\bibnamefont {Shapourian}},
  \bibinfo {author} {\bibfnamefont {A.}~\bibnamefont {Vishwanath}}, \ and\
  \bibinfo {author} {\bibfnamefont {M.~A.}\ \bibnamefont {Metlitski}},\
  }\bibfield  {title} {\enquote {\bibinfo {title} {Magnetic impurities at
  quantum critical points: Large-$n$ expansion and connections to
  symmetry-protected topological states},}\ }\href {\doibase
  10.1103/PhysRevB.104.104201} {\bibfield  {journal} {\bibinfo  {journal}
  {Phys. Rev. B}\ }\textbf {\bibinfo {volume} {104}},\ \bibinfo {pages}
  {104201} (\bibinfo {year} {2021})},\ \Eprint
  {http://arxiv.org/abs/2104.15026} {arXiv:2104.15026 [cond-mat]} \BibitemShut
  {NoStop}%
\bibitem [{\citenamefont {Dantchev}\ and\ \citenamefont
  {Dietrich}()}]{dantchev2022critical}%
  \BibitemOpen
  \bibfield  {author} {\bibinfo {author} {\bibfnamefont {D.~M.}\ \bibnamefont
  {Dantchev}}\ and\ \bibinfo {author} {\bibfnamefont {S.}~\bibnamefont
  {Dietrich}},\ }\bibfield  {title} {\enquote {\bibinfo {title} {Critical
  casimir effect: Exact results},}\ }\href@noop {} {\ }\Eprint
  {http://arxiv.org/abs/2203.15050} {arXiv:2203.15050 [cond-mat]} \BibitemShut
  {NoStop}%
\bibitem [{\citenamefont {Andrei}\ \emph {et~al.}(2020)\citenamefont {Andrei},
  \citenamefont {Bissi}, \citenamefont {Buican}, \citenamefont {Cardy},
  \citenamefont {Dorey}, \citenamefont {Drukker}, \citenamefont {Erdmenger},
  \citenamefont {Friedan}, \citenamefont {Fursaev}, \citenamefont {Konechny},
  \citenamefont {Kristjansen}, \citenamefont {Makabe}, \citenamefont
  {Nakayama}, \citenamefont {O'Bannon}, \citenamefont {Parini}, \citenamefont
  {Robinson}, \citenamefont {Ryu}, \citenamefont {Schmidt-Colinet},
  \citenamefont {Schomerus}, \citenamefont {Schweigert},\ and\ \citenamefont
  {Watts}}]{Andrei_2020}%
  \BibitemOpen
  \bibfield  {author} {\bibinfo {author} {\bibfnamefont {N.}~\bibnamefont
  {Andrei}}, \bibinfo {author} {\bibfnamefont {A.}~\bibnamefont {Bissi}},
  \bibinfo {author} {\bibfnamefont {M.}~\bibnamefont {Buican}}, \bibinfo
  {author} {\bibfnamefont {J.}~\bibnamefont {Cardy}}, \bibinfo {author}
  {\bibfnamefont {P.}~\bibnamefont {Dorey}}, \bibinfo {author} {\bibfnamefont
  {N.}~\bibnamefont {Drukker}}, \bibinfo {author} {\bibfnamefont
  {J.}~\bibnamefont {Erdmenger}}, \bibinfo {author} {\bibfnamefont
  {D.}~\bibnamefont {Friedan}}, \bibinfo {author} {\bibfnamefont
  {D.}~\bibnamefont {Fursaev}}, \bibinfo {author} {\bibfnamefont
  {A.}~\bibnamefont {Konechny}}, \bibinfo {author} {\bibfnamefont
  {C.}~\bibnamefont {Kristjansen}}, \bibinfo {author} {\bibfnamefont
  {I.}~\bibnamefont {Makabe}}, \bibinfo {author} {\bibfnamefont
  {Y.}~\bibnamefont {Nakayama}}, \bibinfo {author} {\bibfnamefont
  {A.}~\bibnamefont {O'Bannon}}, \bibinfo {author} {\bibfnamefont
  {R.}~\bibnamefont {Parini}}, \bibinfo {author} {\bibfnamefont
  {B.}~\bibnamefont {Robinson}}, \bibinfo {author} {\bibfnamefont
  {S.}~\bibnamefont {Ryu}}, \bibinfo {author} {\bibfnamefont {C.}~\bibnamefont
  {Schmidt-Colinet}}, \bibinfo {author} {\bibfnamefont {V.}~\bibnamefont
  {Schomerus}}, \bibinfo {author} {\bibfnamefont {C.}~\bibnamefont
  {Schweigert}}, \ and\ \bibinfo {author} {\bibfnamefont {G.~M.~T.}\
  \bibnamefont {Watts}},\ }\bibfield  {title} {\enquote {\bibinfo {title}
  {Boundary and defect {CFT}: open problems and applications},}\ }\href
  {\doibase 10.1088/1751-8121/abb0fe} {\bibfield  {journal} {\bibinfo
  {journal} {J. Phys. A: Math. and Theo.}\ }\textbf {\bibinfo {volume} {53}},\
  \bibinfo {pages} {453002} (\bibinfo {year} {2020})},\ \Eprint
  {http://arxiv.org/abs/1810.05697} {arXiv:1810.05697 [cond-mat]} \BibitemShut
  {NoStop}%
\bibitem [{\citenamefont {Metlitski}(2022)}]{metlitski2020boundary}%
  \BibitemOpen
  \bibfield  {author} {\bibinfo {author} {\bibfnamefont {M.~A.}\ \bibnamefont
  {Metlitski}},\ }\bibfield  {title} {\enquote {\bibinfo {title} {Boundary
  criticality of the o(n) model in d=3 critically revisited},}\ }\href
  {\doibase 10.21468/SciPostPhys.12.4.131} {\bibfield  {journal} {\bibinfo
  {journal} {SciPost Phys.}\ }\textbf {\bibinfo {volume} {12}},\ \bibinfo
  {pages} {131} (\bibinfo {year} {2022})},\ \Eprint
  {http://arxiv.org/abs/2009.05119} {arXiv:2009.05119 [cond-mat]} \BibitemShut
  {NoStop}%
\bibitem [{\citenamefont {Parisen~Toldin}(2021)}]{toldin2020boundary}%
  \BibitemOpen
  \bibfield  {author} {\bibinfo {author} {\bibfnamefont {F.}~\bibnamefont
  {Parisen~Toldin}},\ }\bibfield  {title} {\enquote {\bibinfo {title} {Boundary
  critical behavior of the three-dimensional heisenberg universality class},}\
  }\href {https://journals.aps.org/prl/abstract/10.1103/PhysRevLett.126.135701}
  {\bibfield  {journal} {\bibinfo  {journal} {Phys. Rev. Lett.}\ }\textbf
  {\bibinfo {volume} {126}},\ \bibinfo {pages} {135701} (\bibinfo {year}
  {2021})},\ \Eprint {http://arxiv.org/abs/2012.00039} {arXiv:2012.00039
  [cond-mat]} \BibitemShut {NoStop}%
\bibitem [{\citenamefont {Hu}\ \emph {et~al.}(2021)\citenamefont {Hu},
  \citenamefont {Deng},\ and\ \citenamefont {Lv}}]{Hu2021}%
  \BibitemOpen
  \bibfield  {author} {\bibinfo {author} {\bibfnamefont {M.}~\bibnamefont
  {Hu}}, \bibinfo {author} {\bibfnamefont {Y.}~\bibnamefont {Deng}}, \ and\
  \bibinfo {author} {\bibfnamefont {J.-P.}\ \bibnamefont {Lv}},\ }\bibfield
  {title} {\enquote {\bibinfo {title} {Extraordinary-log surface phase
  transition in the three-dimensional $xy$ model},}\ }\href {\doibase
  10.1103/PhysRevLett.127.120603} {\bibfield  {journal} {\bibinfo  {journal}
  {Phys. Rev. Lett.}\ }\textbf {\bibinfo {volume} {127}},\ \bibinfo {pages}
  {120603} (\bibinfo {year} {2021})},\ \Eprint
  {http://arxiv.org/abs/2104.05152} {arXiv:2104.05152 [cond-mat]} \BibitemShut
  {NoStop}%
\bibitem [{\citenamefont {Padayasi}\ \emph {et~al.}(2022)\citenamefont
  {Padayasi}, \citenamefont {Krishnan}, \citenamefont {Metlitski},
  \citenamefont {Gruzberg},\ and\ \citenamefont
  {Meineri}}]{padayasi2021extraordinary}%
  \BibitemOpen
  \bibfield  {author} {\bibinfo {author} {\bibfnamefont {J.}~\bibnamefont
  {Padayasi}}, \bibinfo {author} {\bibfnamefont {A.}~\bibnamefont {Krishnan}},
  \bibinfo {author} {\bibfnamefont {M.~A.}\ \bibnamefont {Metlitski}}, \bibinfo
  {author} {\bibfnamefont {I.~A.}\ \bibnamefont {Gruzberg}}, \ and\ \bibinfo
  {author} {\bibfnamefont {M.}~\bibnamefont {Meineri}},\ }\bibfield  {title}
  {\enquote {\bibinfo {title} {The extraordinary boundary transition in the 3d
  o(n) model via conformal bootstrap},}\ }\href
  {https://www.scipost.org/SciPostPhys.12.6.190} {\bibfield  {journal}
  {\bibinfo  {journal} {SciPost Phys.}\ }\textbf {\bibinfo {volume} {12}},\
  \bibinfo {pages} {190} (\bibinfo {year} {2022})},\ \Eprint
  {http://arxiv.org/abs/2111.03071} {arXiv:2111.03071 [cond-mat]} \BibitemShut
  {NoStop}%
\bibitem [{\citenamefont {Parisen~Toldin}\ and\ \citenamefont
  {Metlitski}(2022)}]{ToldinMetlitski2021extraordinary}%
  \BibitemOpen
  \bibfield  {author} {\bibinfo {author} {\bibfnamefont {F.}~\bibnamefont
  {Parisen~Toldin}}\ and\ \bibinfo {author} {\bibfnamefont {M.~A.}\
  \bibnamefont {Metlitski}},\ }\bibfield  {title} {\enquote {\bibinfo {title}
  {Boundary criticality of the 3d o($n$) model: From normal to
  extraordinary},}\ }\href {\doibase 10.1103/PhysRevLett.128.215701} {\bibfield
   {journal} {\bibinfo  {journal} {Phys. Rev. Lett.}\ }\textbf {\bibinfo
  {volume} {128}},\ \bibinfo {pages} {215701} (\bibinfo {year} {2022})},\
  \Eprint {http://arxiv.org/abs/2111.03613} {arXiv:2111.03613 [cond-mat]}
  \BibitemShut {NoStop}%
\bibitem [{\citenamefont {Parisen~Toldin}(2022)}]{ToldinSurface2021}%
  \BibitemOpen
  \bibfield  {author} {\bibinfo {author} {\bibfnamefont {F.}~\bibnamefont
  {Parisen~Toldin}},\ }\bibfield  {title} {\enquote {\bibinfo {title} {Surface
  critical behavior of the three-dimensional o(3) model},}\ }\href
  {https://iopscience.iop.org/article/10.1088/1742-6596/2207/1/012003/meta}
  {\bibfield  {journal} {\bibinfo  {journal} {J. Phys.: Conf. Ser.}\ }\textbf
  {\bibinfo {volume} {2207}},\ \bibinfo {pages} {012003} (\bibinfo {year}
  {2022})},\ \Eprint {http://arxiv.org/abs/2111.11762} {arXiv:2111.11762
  [cond-mat]} \BibitemShut {NoStop}%
\bibitem [{\citenamefont {Zhang}\ \emph {et~al.}(2022)\citenamefont {Zhang},
  \citenamefont {Ding}, \citenamefont {Deng},\ and\ \citenamefont
  {Zhang}}]{Zhang2022Surface}%
  \BibitemOpen
  \bibfield  {author} {\bibinfo {author} {\bibfnamefont {L.-R.}\ \bibnamefont
  {Zhang}}, \bibinfo {author} {\bibfnamefont {C.}~\bibnamefont {Ding}},
  \bibinfo {author} {\bibfnamefont {Y.}~\bibnamefont {Deng}}, \ and\ \bibinfo
  {author} {\bibfnamefont {L.}~\bibnamefont {Zhang}},\ }\bibfield  {title}
  {\enquote {\bibinfo {title} {Surface criticality of the antiferromagnetic
  potts model},}\ }\href {\doibase 10.1103/PhysRevB.105.224415} {\bibfield
  {journal} {\bibinfo  {journal} {Phys. Rev. B}\ }\textbf {\bibinfo {volume}
  {105}},\ \bibinfo {pages} {224415} (\bibinfo {year} {2022})},\ \Eprint
  {http://arxiv.org/abs/2204.11692} {arXiv:2204.11692 [cond-mat]} \BibitemShut
  {NoStop}%
\bibitem [{\citenamefont {Zou}\ \emph {et~al.}(2022)\citenamefont {Zou},
  \citenamefont {Liu},\ and\ \citenamefont {Guo}}]{Zou2022Surface}%
  \BibitemOpen
  \bibfield  {author} {\bibinfo {author} {\bibfnamefont {X.}~\bibnamefont
  {Zou}}, \bibinfo {author} {\bibfnamefont {S}~\bibnamefont {Liu}}, \ and\
  \bibinfo {author} {\bibfnamefont {W.}~\bibnamefont {Guo}},\ }\bibfield
  {title} {\enquote {\bibinfo {title} {Surface critical properties of the
  three-dimensional clock model},}\ }\href {\doibase
  10.1103/PhysRevB.106.064420} {\bibfield  {journal} {\bibinfo  {journal}
  {Phys. Rev. B}\ }\textbf {\bibinfo {volume} {106}},\ \bibinfo {pages}
  {064420} (\bibinfo {year} {2022})},\ \Eprint
  {http://arxiv.org/abs/2204.13612} {arXiv:2204.13612 [cond-mat]} \BibitemShut
  {NoStop}%
\bibitem [{\citenamefont {Weber}\ and\ \citenamefont
  {Wessel}(2021)}]{Weber2021}%
  \BibitemOpen
  \bibfield  {author} {\bibinfo {author} {\bibfnamefont {L.}~\bibnamefont
  {Weber}}\ and\ \bibinfo {author} {\bibfnamefont {S.}~\bibnamefont {Wessel}},\
  }\bibfield  {title} {\enquote {\bibinfo {title} {Spin versus bond
  correlations along dangling edges of quantum critical magnets},}\ }\href
  {\doibase 10.1103/PhysRevB.103.L020406} {\bibfield  {journal} {\bibinfo
  {journal} {Phys. Rev. B}\ }\textbf {\bibinfo {volume} {103}},\ \bibinfo
  {pages} {L020406} (\bibinfo {year} {2021})},\ \Eprint
  {http://arxiv.org/abs/2010.15691} {arXiv:2010.15691 [cond-mat]} \BibitemShut
  {NoStop}%
\bibitem [{\citenamefont {Ding}\ \emph {et~al.}()\citenamefont {Ding},
  \citenamefont {Zhu}, \citenamefont {Guo},\ and\ \citenamefont
  {Zhang}}]{ding2021special}%
  \BibitemOpen
  \bibfield  {author} {\bibinfo {author} {\bibfnamefont {C.}~\bibnamefont
  {Ding}}, \bibinfo {author} {\bibfnamefont {W.}~\bibnamefont {Zhu}}, \bibinfo
  {author} {\bibfnamefont {W.}~\bibnamefont {Guo}}, \ and\ \bibinfo {author}
  {\bibfnamefont {L.}~\bibnamefont {Zhang}},\ }\bibfield  {title} {\enquote
  {\bibinfo {title} {Special transition and extraordinary phase on the surface
  of a (2+ 1)-dimensional quantum heisenberg antiferromagnet},}\ }\href@noop {}
  {\ }\Eprint {http://arxiv.org/abs/2110.04762} {arXiv:2110.04762 [cond-mat]}
  \BibitemShut {NoStop}%
\bibitem [{\citenamefont {Zhu}\ \emph {et~al.}()\citenamefont {Zhu},
  \citenamefont {Ding}, \citenamefont {Zhang},\ and\ \citenamefont
  {W.}}]{Zhu2021Exotic}%
  \BibitemOpen
  \bibfield  {author} {\bibinfo {author} {\bibfnamefont {W.}~\bibnamefont
  {Zhu}}, \bibinfo {author} {\bibfnamefont {C.}~\bibnamefont {Ding}}, \bibinfo
  {author} {\bibfnamefont {L.}~\bibnamefont {Zhang}}, \ and\ \bibinfo {author}
  {\bibfnamefont {Guo}\ \bibnamefont {W.}},\ }\bibfield  {title} {\enquote
  {\bibinfo {title} {Exotic surface behaviors induced by geometrical settings
  of two-dimensional dimerized quantum xxz model},}\ }\href@noop {} {\ }\Eprint
  {http://arxiv.org/abs/2111.12336} {arXiv:2111.12336 [cond-mat]} \BibitemShut
  {NoStop}%
\bibitem [{\citenamefont {Yu}\ \emph {et~al.}(2022)\citenamefont {Yu},
  \citenamefont {Huang}, \citenamefont {Song}, \citenamefont {Xu},
  \citenamefont {Ding},\ and\ \citenamefont {Zhang}}]{Yu2021Conformal}%
  \BibitemOpen
  \bibfield  {author} {\bibinfo {author} {\bibfnamefont {X.-J.}\ \bibnamefont
  {Yu}}, \bibinfo {author} {\bibfnamefont {R.-Z.}\ \bibnamefont {Huang}},
  \bibinfo {author} {\bibfnamefont {H.-H.}\ \bibnamefont {Song}}, \bibinfo
  {author} {\bibfnamefont {L}~\bibnamefont {Xu}}, \bibinfo {author}
  {\bibfnamefont {C.}~\bibnamefont {Ding}}, \ and\ \bibinfo {author}
  {\bibfnamefont {L.}~\bibnamefont {Zhang}},\ }\bibfield  {title} {\enquote
  {\bibinfo {title} {Conformal boundary conditions of symmetry-enriched quantum
  critical spin chains},}\ }\href {\doibase 10.1103/PhysRevLett.129.210601}
  {\bibfield  {journal} {\bibinfo  {journal} {Phys. Rev. Lett.}\ }\textbf
  {\bibinfo {volume} {129}},\ \bibinfo {pages} {210601} (\bibinfo {year}
  {2022})},\ \Eprint {http://arxiv.org/abs/2111.10945} {arXiv:2111.10945
  [cond-mat]} \BibitemShut {NoStop}%
\bibitem [{\citenamefont {Xu}\ \emph {et~al.}()\citenamefont {Xu},
  \citenamefont {Xiong},\ and\ \citenamefont {Zhang}}]{Xu2021Persistent}%
  \BibitemOpen
  \bibfield  {author} {\bibinfo {author} {\bibfnamefont {Y.}~\bibnamefont
  {Xu}}, \bibinfo {author} {\bibfnamefont {Z.}~\bibnamefont {Xiong}}, \ and\
  \bibinfo {author} {\bibfnamefont {L.}~\bibnamefont {Zhang}},\ }\bibfield
  {title} {\enquote {\bibinfo {title} {Persistent corner spin mode at the
  quantum critical point of a plaquette heisenberg model},}\ }\href@noop {} {\
  }\Eprint {http://arxiv.org/abs/2112.04616} {arXiv:2112.04616 [cond-mat]}
  \BibitemShut {NoStop}%
\bibitem [{\citenamefont {Wittmann}\ and\ \citenamefont
  {Young}(2014)}]{Wittmann2014}%
  \BibitemOpen
  \bibfield  {author} {\bibinfo {author} {\bibfnamefont {M.}~\bibnamefont
  {Wittmann}}\ and\ \bibinfo {author} {\bibfnamefont {A.~P.}\ \bibnamefont
  {Young}},\ }\bibfield  {title} {\enquote {\bibinfo {title} {Finite-size
  scaling above the upper critical dimension},}\ }\href {\doibase
  10.1103/PhysRevE.90.062137} {\bibfield  {journal} {\bibinfo  {journal} {Phys.
  Rev. E}\ }\textbf {\bibinfo {volume} {90}},\ \bibinfo {pages} {062137}
  (\bibinfo {year} {2014})},\ \Eprint {http://arxiv.org/abs/1410.5296}
  {arXiv:1410.5296 [cond-mat]} \BibitemShut {NoStop}%
\bibitem [{\citenamefont {Flores-Sola}\ \emph {et~al.}(2016)\citenamefont
  {Flores-Sola}, \citenamefont {Berche}, \citenamefont {Kenna},\ and\
  \citenamefont {Weigel}}]{Flores-Sola2016}%
  \BibitemOpen
  \bibfield  {author} {\bibinfo {author} {\bibfnamefont {E.}~\bibnamefont
  {Flores-Sola}}, \bibinfo {author} {\bibfnamefont {B.}~\bibnamefont {Berche}},
  \bibinfo {author} {\bibfnamefont {R.}~\bibnamefont {Kenna}}, \ and\ \bibinfo
  {author} {\bibfnamefont {M.}~\bibnamefont {Weigel}},\ }\bibfield  {title}
  {\enquote {\bibinfo {title} {Role of fourier modes in finite-size scaling
  above the upper critical dimension},}\ }\href {\doibase
  10.1103/PhysRevLett.116.115701} {\bibfield  {journal} {\bibinfo  {journal}
  {Phys. Rev. Lett.}\ }\textbf {\bibinfo {volume} {116}},\ \bibinfo {pages}
  {115701} (\bibinfo {year} {2016})},\ \Eprint
  {http://arxiv.org/abs/1511.04321} {arXiv:1511.04321 [cond-mat]} \BibitemShut
  {NoStop}%
\bibitem [{\citenamefont {Papathanakos}(2006)}]{papathanakos2006finite}%
  \BibitemOpen
  \bibfield  {author} {\bibinfo {author} {\bibfnamefont {V.}~\bibnamefont
  {Papathanakos}},\ }\href@noop {} {\emph {\bibinfo {title} {Finite-size
  effects in high-dimensional statistical mechanical systems: The Ising model
  with periodic boundary conditions}}}\ (\bibinfo  {publisher} {Ph.D. thesis,
  Princeton University, Princeton, New Jersey},\ \bibinfo {year}
  {2006})\BibitemShut {NoStop}%
\bibitem [{\citenamefont {Grimm}\ \emph {et~al.}(2017)\citenamefont {Grimm},
  \citenamefont {El{\c{c}}i}, \citenamefont {Zhou}, \citenamefont {Garoni},\
  and\ \citenamefont {Deng}}]{grimm2017geometric}%
  \BibitemOpen
  \bibfield  {author} {\bibinfo {author} {\bibfnamefont {J.}~\bibnamefont
  {Grimm}}, \bibinfo {author} {\bibfnamefont {E.~M.}\ \bibnamefont
  {El{\c{c}}i}}, \bibinfo {author} {\bibfnamefont {Z.}~\bibnamefont {Zhou}},
  \bibinfo {author} {\bibfnamefont {T.~M.}\ \bibnamefont {Garoni}}, \ and\
  \bibinfo {author} {\bibfnamefont {Y.}~\bibnamefont {Deng}},\ }\bibfield
  {title} {\enquote {\bibinfo {title} {Geometric explanation of anomalous
  finite-size scaling in high dimensions},}\ }\href
  {https://journals.aps.org/prl/abstract/10.1103/PhysRevLett.118.115701}
  {\bibfield  {journal} {\bibinfo  {journal} {Phys. Rev. Lett.}\ }\textbf
  {\bibinfo {volume} {118}},\ \bibinfo {pages} {115701} (\bibinfo {year}
  {2017})},\ \Eprint {http://arxiv.org/abs/1612.01722} {arXiv:1612.01722
  [cond-mat]} \BibitemShut {NoStop}%
\bibitem [{\citenamefont {Zhou}\ \emph {et~al.}(2018)\citenamefont {Zhou},
  \citenamefont {Grimm}, \citenamefont {Fang}, \citenamefont {Deng},\ and\
  \citenamefont {Garoni}}]{zhou2018random}%
  \BibitemOpen
  \bibfield  {author} {\bibinfo {author} {\bibfnamefont {Z.}~\bibnamefont
  {Zhou}}, \bibinfo {author} {\bibfnamefont {J.}~\bibnamefont {Grimm}},
  \bibinfo {author} {\bibfnamefont {S.}~\bibnamefont {Fang}}, \bibinfo {author}
  {\bibfnamefont {Y.}~\bibnamefont {Deng}}, \ and\ \bibinfo {author}
  {\bibfnamefont {T.~M.}\ \bibnamefont {Garoni}},\ }\bibfield  {title}
  {\enquote {\bibinfo {title} {Random-length random walks and finite-size
  scaling in high dimensions},}\ }\href
  {https://journals.aps.org/prl/abstract/10.1103/PhysRevLett.121.185701}
  {\bibfield  {journal} {\bibinfo  {journal} {Phys. Rev. Lett.}\ }\textbf
  {\bibinfo {volume} {121}},\ \bibinfo {pages} {185701} (\bibinfo {year}
  {2018})},\ \Eprint {http://arxiv.org/abs/1809.00515} {arXiv:1809.00515
  [cond-mat]} \BibitemShut {NoStop}%
\bibitem [{\citenamefont {Fang}\ \emph {et~al.}(2020)\citenamefont {Fang},
  \citenamefont {Grimm}, \citenamefont {Zhou},\ and\ \citenamefont
  {Deng}}]{FangComplete}%
  \BibitemOpen
  \bibfield  {author} {\bibinfo {author} {\bibfnamefont {S.}~\bibnamefont
  {Fang}}, \bibinfo {author} {\bibfnamefont {J.}~\bibnamefont {Grimm}},
  \bibinfo {author} {\bibfnamefont {Z.}~\bibnamefont {Zhou}}, \ and\ \bibinfo
  {author} {\bibfnamefont {Y.}~\bibnamefont {Deng}},\ }\bibfield  {title}
  {\enquote {\bibinfo {title} {Complete graph and gaussian fixed-point
  asymptotics in the five-dimensional fortuin-kasteleyn ising model with
  periodic boundaries},}\ }\href {\doibase 10.1103/PhysRevE.102.022125}
  {\bibfield  {journal} {\bibinfo  {journal} {Phys. Rev. E}\ }\textbf {\bibinfo
  {volume} {102}},\ \bibinfo {pages} {022125} (\bibinfo {year} {2020})},\
  \Eprint {http://arxiv.org/abs/1909.04328} {arXiv:1909.04328 [cond-mat]}
  \BibitemShut {NoStop}%
\bibitem [{\citenamefont {Lv}\ \emph {et~al.}(2021)\citenamefont {Lv},
  \citenamefont {Xu}, \citenamefont {Sun}, \citenamefont {Chen},\ and\
  \citenamefont {Deng}}]{lv2021finite}%
  \BibitemOpen
  \bibfield  {author} {\bibinfo {author} {\bibfnamefont {J.-P.}\ \bibnamefont
  {Lv}}, \bibinfo {author} {\bibfnamefont {W.}~\bibnamefont {Xu}}, \bibinfo
  {author} {\bibfnamefont {Y.}~\bibnamefont {Sun}}, \bibinfo {author}
  {\bibfnamefont {K.}~\bibnamefont {Chen}}, \ and\ \bibinfo {author}
  {\bibfnamefont {Y.}~\bibnamefont {Deng}},\ }\bibfield  {title} {\enquote
  {\bibinfo {title} {Finite-size scaling of o(n) systems at the upper critical
  dimensionality},}\ }\href
  {https://academic.oup.com/nsr/article/8/3/nwaa212/5899772?login=true}
  {\bibfield  {journal} {\bibinfo  {journal} {Natl. Sci. Rev.}\ }\textbf
  {\bibinfo {volume} {8}},\ \bibinfo {pages} {nwaa212} (\bibinfo {year}
  {2021})},\ \Eprint {http://arxiv.org/abs/1909.10347} {arXiv:1909.10347
  [cond-mat]} \BibitemShut {NoStop}%
\bibitem [{\citenamefont {Fang}\ \emph {et~al.}(2021)\citenamefont {Fang},
  \citenamefont {Deng},\ and\ \citenamefont {Zhou}}]{Fang2021}%
  \BibitemOpen
  \bibfield  {author} {\bibinfo {author} {\bibfnamefont {S.}~\bibnamefont
  {Fang}}, \bibinfo {author} {\bibfnamefont {Y.}~\bibnamefont {Deng}}, \ and\
  \bibinfo {author} {\bibfnamefont {Z.}~\bibnamefont {Zhou}},\ }\bibfield
  {title} {\enquote {\bibinfo {title} {Logarithmic finite-size scaling of the
  self-avoiding walk at four dimensions},}\ }\href {\doibase
  10.1103/PhysRevE.104.064108} {\bibfield  {journal} {\bibinfo  {journal}
  {Phys. Rev. E}\ }\textbf {\bibinfo {volume} {104}},\ \bibinfo {pages}
  {064108} (\bibinfo {year} {2021})},\ \Eprint
  {http://arxiv.org/abs/2103.04340} {arXiv:2103.04340 [cond-mat]} \BibitemShut
  {NoStop}%
\bibitem [{\citenamefont {Shao}\ \emph {et~al.}(2016)\citenamefont {Shao},
  \citenamefont {Guo},\ and\ \citenamefont {Sandvik}}]{shao2016quantum}%
  \BibitemOpen
  \bibfield  {author} {\bibinfo {author} {\bibfnamefont {H.}~\bibnamefont
  {Shao}}, \bibinfo {author} {\bibfnamefont {W.}~\bibnamefont {Guo}}, \ and\
  \bibinfo {author} {\bibfnamefont {A.~W.}\ \bibnamefont {Sandvik}},\
  }\bibfield  {title} {\enquote {\bibinfo {title} {Quantum criticality with two
  length scales},}\ }\href
  {https://science.sciencemag.org/content/352/6282/213.abstract} {\bibfield
  {journal} {\bibinfo  {journal} {Science}\ }\textbf {\bibinfo {volume}
  {352}},\ \bibinfo {pages} {213--216} (\bibinfo {year} {2016})},\ \Eprint
  {http://arxiv.org/abs/1603.02171} {arXiv:1603.02171 [cond-mat]} \BibitemShut
  {NoStop}%
\bibitem [{\citenamefont {Heydenreich}\ and\ \citenamefont {Van~der
  Hofstad}(2017)}]{heydenreich2017progress}%
  \BibitemOpen
  \bibfield  {author} {\bibinfo {author} {\bibfnamefont {M.}~\bibnamefont
  {Heydenreich}}\ and\ \bibinfo {author} {\bibfnamefont {R.}~\bibnamefont
  {Van~der Hofstad}},\ }\href
  {https://link.springer.com/chapter/10.1007/978-3-319-62473-0_13} {\emph
  {\bibinfo {title} {Progress in high-dimensional percolation and random
  graphs}}}\ (\bibinfo  {publisher} {Springer},\ \bibinfo {year}
  {2017})\BibitemShut {NoStop}%
\bibitem [{\citenamefont {Bet}\ \emph {et~al.}(2021)\citenamefont {Bet},
  \citenamefont {Bogerd}, \citenamefont {Castro},\ and\ \citenamefont {van~der
  Hofstad}}]{bet2021detecting}%
  \BibitemOpen
  \bibfield  {author} {\bibinfo {author} {\bibfnamefont {G.}~\bibnamefont
  {Bet}}, \bibinfo {author} {\bibfnamefont {K.}~\bibnamefont {Bogerd}},
  \bibinfo {author} {\bibfnamefont {R.~M.}\ \bibnamefont {Castro}}, \ and\
  \bibinfo {author} {\bibfnamefont {R.}~\bibnamefont {van~der Hofstad}},\
  }\bibfield  {title} {\enquote {\bibinfo {title} {Detecting a botnet in a
  network},}\ }\href {https://ems.press/journals/msl/articles/3922907}
  {\bibfield  {journal} {\bibinfo  {journal} {Math. Stat. Learn.}\ }\textbf
  {\bibinfo {volume} {3}},\ \bibinfo {pages} {315--343} (\bibinfo {year}
  {2021})},\ \Eprint {http://arxiv.org/abs/2005.10650} {arXiv:2005.10650
  [cond-mat]} \BibitemShut {NoStop}%
\bibitem [{\citenamefont {Deng}\ \emph {et~al.}(2022)\citenamefont {Deng},
  \citenamefont {Garoni}, \citenamefont {Grimm},\ and\ \citenamefont
  {Zhou}}]{deng2022unwrapped}%
  \BibitemOpen
  \bibfield  {author} {\bibinfo {author} {\bibfnamefont {Y.}~\bibnamefont
  {Deng}}, \bibinfo {author} {\bibfnamefont {T.~M.}\ \bibnamefont {Garoni}},
  \bibinfo {author} {\bibfnamefont {J.}~\bibnamefont {Grimm}}, \ and\ \bibinfo
  {author} {\bibfnamefont {Z.}~\bibnamefont {Zhou}},\ }\bibfield  {title}
  {\enquote {\bibinfo {title} {Unwrapped two-point functions on
  high-dimensional tori},}\ }\href
  {https://iopscience.iop.org/article/10.1088/1742-5468/ac6a5c/meta} {\bibfield
   {journal} {\bibinfo  {journal} {J. Stat. Mech.: Theo. and Exp.}\ }\textbf
  {\bibinfo {volume} {2022}},\ \bibinfo {pages} {053208} (\bibinfo {year}
  {2022})},\ \Eprint {http://arxiv.org/abs/2203.05100} {arXiv:2203.05100
  [cond-mat]} \BibitemShut {NoStop}%
\bibitem [{\citenamefont {Jian}\ \emph {et~al.}(2021)\citenamefont {Jian},
  \citenamefont {Xu}, \citenamefont {Wu},\ and\ \citenamefont
  {Xu}}]{jian2021continuous}%
  \BibitemOpen
  \bibfield  {author} {\bibinfo {author} {\bibfnamefont {C.-M.}\ \bibnamefont
  {Jian}}, \bibinfo {author} {\bibfnamefont {Y.}~\bibnamefont {Xu}}, \bibinfo
  {author} {\bibfnamefont {X.-C.}\ \bibnamefont {Wu}}, \ and\ \bibinfo {author}
  {\bibfnamefont {C.}~\bibnamefont {Xu}},\ }\bibfield  {title} {\enquote
  {\bibinfo {title} {Continuous neel-vbs quantum phase transition in non-local
  one-dimensional systems with so(3) symmetry},}\ }\href
  {https://www.scipost.org/SciPostPhys.10.2.033} {\bibfield  {journal}
  {\bibinfo  {journal} {SciPost Physics}\ }\textbf {\bibinfo {volume} {10}},\
  \bibinfo {pages} {033} (\bibinfo {year} {2021})},\ \Eprint
  {http://arxiv.org/abs/2004.07852} {arXiv:2004.07852 [cond-mat]} \BibitemShut
  {NoStop}%
\bibitem [{\citenamefont {Fisher}\ \emph {et~al.}(1989)\citenamefont {Fisher},
  \citenamefont {Weichman}, \citenamefont {Grinstein},\ and\ \citenamefont
  {Fisher}}]{Fisher1989}%
  \BibitemOpen
  \bibfield  {author} {\bibinfo {author} {\bibfnamefont {M.~P.~A.}\
  \bibnamefont {Fisher}}, \bibinfo {author} {\bibfnamefont {P.~B.}\
  \bibnamefont {Weichman}}, \bibinfo {author} {\bibfnamefont {G.}~\bibnamefont
  {Grinstein}}, \ and\ \bibinfo {author} {\bibfnamefont {D.~S.}\ \bibnamefont
  {Fisher}},\ }\bibfield  {title} {\enquote {\bibinfo {title} {Boson
  localization and the superfluid-insulator transition},}\ }\href {\doibase
  10.1103/PhysRevB.40.546} {\bibfield  {journal} {\bibinfo  {journal} {Phys.
  Rev. B}\ }\textbf {\bibinfo {volume} {40}},\ \bibinfo {pages} {546--570}
  (\bibinfo {year} {1989})}\BibitemShut {NoStop}%
\bibitem [{\citenamefont {Xu}\ \emph {et~al.}(2019)\citenamefont {Xu},
  \citenamefont {Sun}, \citenamefont {Lv},\ and\ \citenamefont
  {Deng}}]{Xu2019}%
  \BibitemOpen
  \bibfield  {author} {\bibinfo {author} {\bibfnamefont {W.}~\bibnamefont
  {Xu}}, \bibinfo {author} {\bibfnamefont {Y.}~\bibnamefont {Sun}}, \bibinfo
  {author} {\bibfnamefont {J.-P.}\ \bibnamefont {Lv}}, \ and\ \bibinfo {author}
  {\bibfnamefont {Y.}~\bibnamefont {Deng}},\ }\bibfield  {title} {\enquote
  {\bibinfo {title} {High-precision monte carlo study of several models in the
  three-dimensional u(1) universality class},}\ }\href {\doibase
  10.1103/PhysRevB.100.064525} {\bibfield  {journal} {\bibinfo  {journal}
  {Phys. Rev. B}\ }\textbf {\bibinfo {volume} {100}},\ \bibinfo {pages}
  {064525} (\bibinfo {year} {2019})},\ \Eprint
  {http://arxiv.org/abs/1908.10990} {arXiv:1908.10990 [cond-mat]} \BibitemShut
  {NoStop}%
\bibitem [{\citenamefont {Capogrosso-Sansone}\ \emph
  {et~al.}(2008)\citenamefont {Capogrosso-Sansone}, \citenamefont {S\"oyler},
  \citenamefont {Prokof'ev},\ and\ \citenamefont
  {Svistunov}}]{capogrosso2008monte}%
  \BibitemOpen
  \bibfield  {author} {\bibinfo {author} {\bibfnamefont {B.}~\bibnamefont
  {Capogrosso-Sansone}}, \bibinfo {author} {\bibfnamefont {{\c{S}}.~G.}\
  \bibnamefont {S\"oyler}}, \bibinfo {author} {\bibfnamefont {N.}~\bibnamefont
  {Prokof'ev}}, \ and\ \bibinfo {author} {\bibfnamefont {B.}~\bibnamefont
  {Svistunov}},\ }\bibfield  {title} {\enquote {\bibinfo {title} {Monte carlo
  study of the two-dimensional bose-hubbard model},}\ }\href {\doibase
  10.1103/PhysRevA.77.015602} {\bibfield  {journal} {\bibinfo  {journal} {Phys.
  Rev. A}\ }\textbf {\bibinfo {volume} {77}},\ \bibinfo {pages} {015602}
  (\bibinfo {year} {2008})},\ \Eprint {http://arxiv.org/abs/0710.2703}
  {arXiv:0710.2703 [cond-mat]} \BibitemShut {NoStop}%
\bibitem [{Note1()}]{Note1}%
  \BibitemOpen
  \bibinfo {note} {Logarithmic corrections may emerge for the KT-like
  transition and the SE-MIB phase~\cite {kosterlitz2016kosterlitz}}\BibitemShut
  {NoStop}%
\bibitem [{\citenamefont {Prokof'ev}\ \emph
  {et~al.}(1998{\natexlab{a}})\citenamefont {Prokof'ev}, \citenamefont
  {Svistunov},\ and\ \citenamefont {Tupitsyn}}]{prokofev1998exact}%
  \BibitemOpen
  \bibfield  {author} {\bibinfo {author} {\bibfnamefont {N.~V.}\ \bibnamefont
  {Prokof'ev}}, \bibinfo {author} {\bibfnamefont {B.~V.}\ \bibnamefont
  {Svistunov}}, \ and\ \bibinfo {author} {\bibfnamefont {I.~S.}\ \bibnamefont
  {Tupitsyn}},\ }\bibfield  {title} {\enquote {\bibinfo {title} {Exact,
  complete, and universal continuous-time worldline monte carlo approach to the
  statistics of discrete quantum systems},}\ }\href
  {https://link.springer.com/article/10.1134/1.558661} {\bibfield  {journal}
  {\bibinfo  {journal} {Sov. Phys. JETP}\ }\textbf {\bibinfo {volume} {87}},\
  \bibinfo {pages} {310--321} (\bibinfo {year} {1998}{\natexlab{a}})},\ \Eprint
  {http://arxiv.org/abs/cond-mat/9703200} {arXiv:cond-mat/9703200 [cond-mat]}
  \BibitemShut {NoStop}%
\bibitem [{\citenamefont {Prokof'ev}\ \emph
  {et~al.}(1998{\natexlab{b}})\citenamefont {Prokof'ev}, \citenamefont
  {Svistunov},\ and\ \citenamefont {Tupitsyn}}]{prokofev1998worm}%
  \BibitemOpen
  \bibfield  {author} {\bibinfo {author} {\bibfnamefont {N.~V.}\ \bibnamefont
  {Prokof'ev}}, \bibinfo {author} {\bibfnamefont {B.~V.}\ \bibnamefont
  {Svistunov}}, \ and\ \bibinfo {author} {\bibfnamefont {I.~S.}\ \bibnamefont
  {Tupitsyn}},\ }\bibfield  {title} {\enquote {\bibinfo {title} {``worm''
  algorithm in quantum monte carlo simulations},}\ }\href
  {https://www.sciencedirect.com/science/article/abs/pii/S0375960197009572}
  {\bibfield  {journal} {\bibinfo  {journal} {Phys. Lett. A}\ }\textbf
  {\bibinfo {volume} {238}},\ \bibinfo {pages} {253--257} (\bibinfo {year}
  {1998}{\natexlab{b}})}\BibitemShut {NoStop}%
\bibitem [{\citenamefont {Grassberger}(2003)}]{Grassberger2003}%
  \BibitemOpen
  \bibfield  {author} {\bibinfo {author} {\bibfnamefont {P.}~\bibnamefont
  {Grassberger}},\ }\bibfield  {title} {\enquote {\bibinfo {title} {Critical
  percolation in high dimensions},}\ }\href {\doibase
  10.1103/PhysRevE.67.036101} {\bibfield  {journal} {\bibinfo  {journal} {Phys.
  Rev. E}\ }\textbf {\bibinfo {volume} {67}},\ \bibinfo {pages} {036101}
  (\bibinfo {year} {2003})},\ \Eprint {http://arxiv.org/abs/cond-mat/0202144}
  {arXiv:cond-mat/0202144 [cond-mat]} \BibitemShut {NoStop}%
\bibitem [{\citenamefont {Guida}\ and\ \citenamefont
  {Zinn-Justin}(1998)}]{guida1998critical}%
  \BibitemOpen
  \bibfield  {author} {\bibinfo {author} {\bibfnamefont {R.}~\bibnamefont
  {Guida}}\ and\ \bibinfo {author} {\bibfnamefont {J.}~\bibnamefont
  {Zinn-Justin}},\ }\bibfield  {title} {\enquote {\bibinfo {title} {Critical
  exponents of the n-vector model},}\ }\href
  {https://iopscience.iop.org/article/10.1088/0305-4470/31/40/006/meta}
  {\bibfield  {journal} {\bibinfo  {journal} {J. Phys. A: Math. Gen.}\ }\textbf
  {\bibinfo {volume} {31}},\ \bibinfo {pages} {8103} (\bibinfo {year}
  {1998})},\ \Eprint {http://arxiv.org/abs/cond-mat/9803240}
  {arXiv:cond-mat/9803240 [cond-mat]} \BibitemShut {NoStop}%
\bibitem [{\citenamefont {Pollock}\ and\ \citenamefont
  {Ceperley}(1987)}]{Pollock1987}%
  \BibitemOpen
  \bibfield  {author} {\bibinfo {author} {\bibfnamefont {E.~L.}\ \bibnamefont
  {Pollock}}\ and\ \bibinfo {author} {\bibfnamefont {D.~M.}\ \bibnamefont
  {Ceperley}},\ }\bibfield  {title} {\enquote {\bibinfo {title} {Path-integral
  computation of superfluid densities},}\ }\href {\doibase
  10.1103/PhysRevB.36.8343} {\bibfield  {journal} {\bibinfo  {journal} {Phys.
  Rev. B}\ }\textbf {\bibinfo {volume} {36}},\ \bibinfo {pages} {8343--8352}
  (\bibinfo {year} {1987})}\BibitemShut {NoStop}%
\bibitem [{\citenamefont {Sun}\ \emph {et~al.}(2022)\citenamefont {Sun},
  \citenamefont {Lyu},\ and\ \citenamefont {Lv}}]{unpublished}%
  \BibitemOpen
  \bibfield  {author} {\bibinfo {author} {\bibfnamefont {Y.}~\bibnamefont
  {Sun}}, \bibinfo {author} {\bibfnamefont {J.}~\bibnamefont {Lyu}}, \ and\
  \bibinfo {author} {\bibfnamefont {J.-P.}\ \bibnamefont {Lv}},\ }\bibfield
  {title} {\enquote {\bibinfo {title} {Classical-quantum correspondence of
  special and extraordinary-log criticality: Villain's bridge},}\ }\href
  {\doibase 10.1103/PhysRevB.106.174516} {\bibfield  {journal} {\bibinfo
  {journal} {Phys. Rev. B}\ }\textbf {\bibinfo {volume} {106}},\ \bibinfo
  {pages} {174516} (\bibinfo {year} {2022})},\ \Eprint
  {http://arxiv.org/abs/2211.11376} {arXiv:2211.11376 [cond-mat]} \BibitemShut
  {NoStop}%
\bibitem [{\citenamefont {Kosterlitz}(1974)}]{kosterlitz1974critical}%
  \BibitemOpen
  \bibfield  {author} {\bibinfo {author} {\bibfnamefont {J.~M.}\ \bibnamefont
  {Kosterlitz}},\ }\bibfield  {title} {\enquote {\bibinfo {title} {The critical
  properties of the two-dimensional xy model},}\ }\href
  {https://iopscience.iop.org/article/10.1088/0022-3719/7/6/005/meta}
  {\bibfield  {journal} {\bibinfo  {journal} {J. Phys. C: Solid State Phys.}\
  }\textbf {\bibinfo {volume} {7}},\ \bibinfo {pages} {1046} (\bibinfo {year}
  {1974})}\BibitemShut {NoStop}%
\bibitem [{\citenamefont {Kosterlitz}(2016)}]{kosterlitz2016kosterlitz}%
  \BibitemOpen
  \bibfield  {author} {\bibinfo {author} {\bibfnamefont {J.~M}\ \bibnamefont
  {Kosterlitz}},\ }\bibfield  {title} {\enquote {\bibinfo {title}
  {Kosterlitz--thouless physics: a review of key issues},}\ }\href
  {https://iopscience.iop.org/article/10.1088/0034-4885/79/2/026001/meta}
  {\bibfield  {journal} {\bibinfo  {journal} {Rep. Prog. Phys.}\ }\textbf
  {\bibinfo {volume} {79}},\ \bibinfo {pages} {026001} (\bibinfo {year}
  {2016})}\BibitemShut {NoStop}%
\bibitem [{\citenamefont {Jaksch}\ \emph {et~al.}(1998)\citenamefont {Jaksch},
  \citenamefont {Bruder}, \citenamefont {Cirac}, \citenamefont {Gardiner},\
  and\ \citenamefont {Zoller}}]{jaksch1998cold}%
  \BibitemOpen
  \bibfield  {author} {\bibinfo {author} {\bibfnamefont {D.}~\bibnamefont
  {Jaksch}}, \bibinfo {author} {\bibfnamefont {C.}~\bibnamefont {Bruder}},
  \bibinfo {author} {\bibfnamefont {J.~I.}\ \bibnamefont {Cirac}}, \bibinfo
  {author} {\bibfnamefont {C.~W.}\ \bibnamefont {Gardiner}}, \ and\ \bibinfo
  {author} {\bibfnamefont {P.}~\bibnamefont {Zoller}},\ }\bibfield  {title}
  {\enquote {\bibinfo {title} {Cold bosonic atoms in optical lattices},}\
  }\href {https://journals.aps.org/prl/abstract/10.1103/PhysRevLett.81.3108}
  {\bibfield  {journal} {\bibinfo  {journal} {Phys. Rev. Lett.}\ }\textbf
  {\bibinfo {volume} {81}},\ \bibinfo {pages} {3108} (\bibinfo {year}
  {1998})},\ \Eprint {http://arxiv.org/abs/cond-mat/9805329}
  {arXiv:cond-mat/9805329 [cond-mat]} \BibitemShut {NoStop}%
\bibitem [{\citenamefont {Greiner}\ \emph {et~al.}(2002)\citenamefont
  {Greiner}, \citenamefont {Mandel}, \citenamefont {Esslinger}, \citenamefont
  {H{\"a}nsch},\ and\ \citenamefont {Bloch}}]{greiner2002quantum}%
  \BibitemOpen
  \bibfield  {author} {\bibinfo {author} {\bibfnamefont {M.}~\bibnamefont
  {Greiner}}, \bibinfo {author} {\bibfnamefont {O.}~\bibnamefont {Mandel}},
  \bibinfo {author} {\bibfnamefont {T.}~\bibnamefont {Esslinger}}, \bibinfo
  {author} {\bibfnamefont {T.~W.}\ \bibnamefont {H{\"a}nsch}}, \ and\ \bibinfo
  {author} {\bibfnamefont {I.}~\bibnamefont {Bloch}},\ }\bibfield  {title}
  {\enquote {\bibinfo {title} {Quantum phase transition from a superfluid to a
  mott insulator in a gas of ultracold atoms},}\ }\href
  {https://www.nature.com/articles/415039a} {\bibfield  {journal} {\bibinfo
  {journal} {Nature}\ }\textbf {\bibinfo {volume} {415}},\ \bibinfo {pages}
  {39--44} (\bibinfo {year} {2002})}\BibitemShut {NoStop}%
\bibitem [{\citenamefont {Baier}\ \emph {et~al.}(2016)\citenamefont {Baier},
  \citenamefont {Mark}, \citenamefont {Petter}, \citenamefont {Aikawa},
  \citenamefont {Chomaz}, \citenamefont {Cai}, \citenamefont {Baranov},
  \citenamefont {Zoller},\ and\ \citenamefont {Ferlaino}}]{baier2016extended}%
  \BibitemOpen
  \bibfield  {author} {\bibinfo {author} {\bibfnamefont {S.}~\bibnamefont
  {Baier}}, \bibinfo {author} {\bibfnamefont {M.~J.}\ \bibnamefont {Mark}},
  \bibinfo {author} {\bibfnamefont {D.}~\bibnamefont {Petter}}, \bibinfo
  {author} {\bibfnamefont {K.}~\bibnamefont {Aikawa}}, \bibinfo {author}
  {\bibfnamefont {L.}~\bibnamefont {Chomaz}}, \bibinfo {author} {\bibfnamefont
  {Z.}~\bibnamefont {Cai}}, \bibinfo {author} {\bibfnamefont {M.}~\bibnamefont
  {Baranov}}, \bibinfo {author} {\bibfnamefont {P.}~\bibnamefont {Zoller}}, \
  and\ \bibinfo {author} {\bibfnamefont {F.}~\bibnamefont {Ferlaino}},\
  }\bibfield  {title} {\enquote {\bibinfo {title} {Extended bose-hubbard models
  with ultracold magnetic atoms},}\ }\href
  {https://www.science.org/doi/abs/10.1126/science.aac9812} {\bibfield
  {journal} {\bibinfo  {journal} {Science}\ }\textbf {\bibinfo {volume}
  {352}},\ \bibinfo {pages} {201--205} (\bibinfo {year} {2016})},\ \Eprint
  {http://arxiv.org/abs/1507.03500} {arXiv:1507.03500 [cond-mat]} \BibitemShut
  {NoStop}%
\bibitem [{\citenamefont {Yang}\ \emph {et~al.}(2020)\citenamefont {Yang},
  \citenamefont {Sun}, \citenamefont {Huang}, \citenamefont {Wang},
  \citenamefont {Deng}, \citenamefont {Dai}, \citenamefont {Yuan},\ and\
  \citenamefont {Pan}}]{yang2020cooling}%
  \BibitemOpen
  \bibfield  {author} {\bibinfo {author} {\bibfnamefont {B.}~\bibnamefont
  {Yang}}, \bibinfo {author} {\bibfnamefont {H.}~\bibnamefont {Sun}}, \bibinfo
  {author} {\bibfnamefont {C.-J.}\ \bibnamefont {Huang}}, \bibinfo {author}
  {\bibfnamefont {H.-Y.}\ \bibnamefont {Wang}}, \bibinfo {author}
  {\bibfnamefont {Y.}~\bibnamefont {Deng}}, \bibinfo {author} {\bibfnamefont
  {H.-N.}\ \bibnamefont {Dai}}, \bibinfo {author} {\bibfnamefont {Z.-S.}\
  \bibnamefont {Yuan}}, \ and\ \bibinfo {author} {\bibfnamefont {J.-W.}\
  \bibnamefont {Pan}},\ }\bibfield  {title} {\enquote {\bibinfo {title}
  {Cooling and entangling ultracold atoms in optical lattices},}\ }\href
  {https://www.science.org/doi/abs/10.1126/science.aaz6801} {\bibfield
  {journal} {\bibinfo  {journal} {Science}\ }\textbf {\bibinfo {volume}
  {369}},\ \bibinfo {pages} {550--553} (\bibinfo {year} {2020})},\ \Eprint
  {http://arxiv.org/abs/1901.01146} {arXiv:1901.01146 [cond-mat]} \BibitemShut
  {NoStop}%
\bibitem [{\citenamefont {Salas}(2020)}]{salas2020phase}%
  \BibitemOpen
  \bibfield  {author} {\bibinfo {author} {\bibfnamefont {J.}~\bibnamefont
  {Salas}},\ }\bibfield  {title} {\enquote {\bibinfo {title} {Phase diagram for
  the bisected-hexagonal-lattice five-state potts antiferromagnet},}\ }\href
  {https://journals.aps.org/pre/abstract/10.1103/PhysRevE.102.032124}
  {\bibfield  {journal} {\bibinfo  {journal} {Phys. Rev. E}\ }\textbf {\bibinfo
  {volume} {102}},\ \bibinfo {pages} {032124} (\bibinfo {year} {2020})},\
  \Eprint {http://arxiv.org/abs/2006.04866} {arXiv:2006.04866 [cond-mat]}
  \BibitemShut {NoStop}%
\bibitem [{\citenamefont {Prokof'ev}\ and\ \citenamefont
  {Svistunov}(2001)}]{prokof2001worm}%
  \BibitemOpen
  \bibfield  {author} {\bibinfo {author} {\bibfnamefont {N.}~\bibnamefont
  {Prokof'ev}}\ and\ \bibinfo {author} {\bibfnamefont {B.}~\bibnamefont
  {Svistunov}},\ }\bibfield  {title} {\enquote {\bibinfo {title} {Worm
  algorithms for classical statistical models},}\ }\href
  {https://journals.aps.org/prl/abstract/10.1103/PhysRevLett.87.160601}
  {\bibfield  {journal} {\bibinfo  {journal} {Phys. Rev. Lett.}\ }\textbf
  {\bibinfo {volume} {87}},\ \bibinfo {pages} {160601} (\bibinfo {year}
  {2001})},\ \Eprint {http://arxiv.org/abs/cond-mat/0103146}
  {arXiv:cond-mat/0103146 [cond-mat]} \BibitemShut {NoStop}%
\end{thebibliography}%

\appendix

\section{Details of methodology}\label{se0}

In the appendixes, we present details for Monte Carlo simulations and provide a benchmark for two-point correlation using bulk criticality. We then analyze the data for the quantum critical phenomena on open edges, which include the special transition, the Kosterlitz-Thouless-like criticality, the ordinary critical phase and the extraordinary-log critical phase.

The raw data are all obtained from quantum Monte Carlo simulations, by means of the worm algorithm in the continuous-time path integral representation. The side lengths of square lattices include $L=16, 32, 48, 64, 96, 128$ and $192$. In the worm simulations, the number of tentative updates for the defects, usually denoted by $Ira$ (I) and $Masha$ (M), ranges from $3.6 \times 10^{12}$ to $3.4 \times 10^{13}$ for $16 \le L \le 48$, and from $1.8 \times 10^{13}$ to $3.7 \times 10^{13}$ for $64 \le L \le 192$.

We perform FSS analyses by using least-squares fits. To this end, we utilize the function {\tt NonlinearModelFit} in {\textsf Mathematica}, as adopted in Ref.~\cite{salas2020phase}. According to standard criterion, we prefer the fits with $\chi^2/{\rm DF} \sim 1$, where $\chi^2/{\rm DF}$ represents the Chi squared per degree of freedom. We draw conclusions by comparing the fits that are stable against varying $L_{\rm min}$, which is the minimum side length incorporated in fitting. In certain situations, we also include a cutoff $L_{\rm max}$ for larger sizes.

\section{Benchmark for two-point correlation using bulk criticality}\label{se1}

We use an estimator of equal-imaginary-time correlations, which avoids reweighting along imaginary-time axis and turns out to be computationally cheap. The estimator correctly captures the asymptotic behavior in the $L \rightarrow \infty$ limit. Specifically speaking, in the worm quantum Monte Carlo simulations, we trace the trajectories of the defects $I$ and $M$ on an edge. If the imaginary-time distance between the defects is less than the $1/L$ fraction of entire axis, the distance $r$ of two defects along the edge is recorded. The follow-up treatment is similar to the measurement of two-point correlations in a classical model~\cite{lv2021finite} which was based on the original idea in Ref.~\cite{prokof2001worm}. We use the $r=1$ result to normalize the two-point correlation and concentrate on the $r \ne 0$ domain of correlation function. Hence, the results do not suffer from the biased allocations of statistical weight between original and Green function state spaces. Finally, we obtain the two-point correlation $g(r)$ as a function of $r$ along the edge.

\begin{figure}
	\includegraphics[height=11cm,width=8cm]{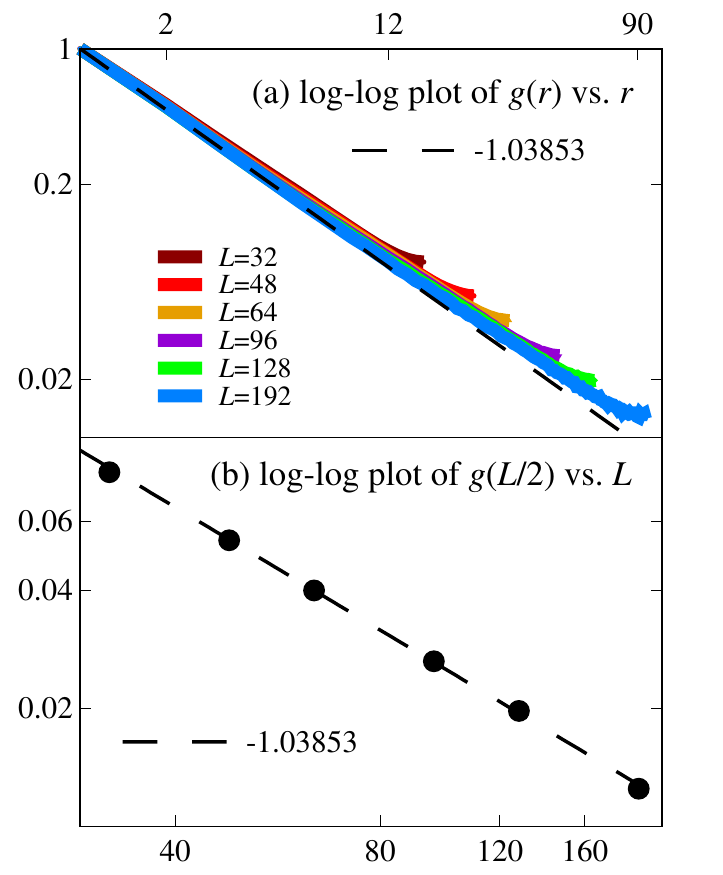}
	\caption{Bulk criticality. (a) Log-log plot of $g(r)$ versus $r$. (b) Log-log plot of $g(L/2)$ versus $L$.}~\label{SM_fig1}
\end{figure}

We proceed to benchmark the above-mentioned methodology for correlation function using the bulk criticality. Particularly, we apply
periodic conditions for both [10] and [01] directions to eliminate the open edges and sample the correlation functions at $t_c$. We analyze the $r$ dependence of $g(r)$ as well as the $L$-dependent behavior of $g(L/2)$. We quote a precise estimate $\eta = 0.03853(48)$ for the anomalous dimension of the (2+1)-dimensional O(2) criticality~\cite{Xu2019}. As shown in Fig.~\ref{SM_fig1}(a), the $r$-dependent behavior converges to the power law $g(r) \sim r^{2-(d+z)-\eta}$, with $d=2$, $z=1$ and $\eta \approx 0.03853$. From Fig.~\ref{SM_fig1}(b), we verify that $g(L/2)$ scales as $g(L/2) \sim L^{-1.03853}$.

More quantitative verification can be achieved by least-squares fits. We fit $g(L/2)$ to
\begin{equation}
g(L/2)=a L^b,
\label{eqfitBOK}
\end{equation}
where $a$ is a constant and $b=-1-\eta$. The results are summarized in Table~\ref{TABLE_B}. We obtain $b=-1.027(6)$ and $\chi^2/{\rm DF} \approx 1.2$ for $L_{\rm min}=48$, $b=-1.03(1)$ and $\chi^2/{\rm DF} \approx 1.5$ for $L_{\rm min}=64$, as well as $b=-1.06(3)$ and $\chi^2/{\rm DF} \approx 1.4$ for $L_{\rm min}=96$. The estimates of $b$ are consistent with $-1-\eta=-1.03853(48)$ of the (2+1)-dimensional O(2) universality.

\section{Details of the FSS analyses for BCB}\label{se2}

In this appendix, we perform FSS analyses for the special transition, the Kosterlitz-Thouless-like criticality, the ordinary critical phase and the extraordinary-log critical phase.

\begin{table}
\begin{center}
\caption{Fits of $g(L/2)$ to Eq.~(\ref{eqfitBOK}) at the bulk critical point.}
\label{TABLE_B}
\begin{tabular}{p{1.2cm}p{1.5cm}p{1.5cm}p{1.3cm}}
\hline
\hline
$L_{\rm min}$ & $\chi^2$/DF &$a$&$b$\\
\hline
32 &20.62/4&2.65(3)&--1.009(3) \\
48  &3.45/3&2.86(6)&--1.027(6)  \\
64  &3.04/2&2.9(1)&--1.03(1) \\
96  &1.41/1&3.4(4)&--1.06(3) \\		
\hline
\hline
\end{tabular}
\end{center}
\end{table}

\headline{Special transition}
We locate the special transition point using the FSS of the winding probability $R_{[10]}$. We perform fits according to
\begin{equation}
	R_{[10]} = R_{[10]}^c+a_1 (\kappa-\kappa_c)L^{y_t}+b_1 L^{-\omega_1},
	\label{eqfitS}
\end{equation}
where $R_{[10]}^c$ is the critical dimensionless ratio, $a_1$ and $b_1$ represent fitting parameters, $\kappa_c$ denotes the transition point, $y_t$ relates to the correlation length exponent $\nu$ by $y_t=1/\nu$, and $\omega_1$ denotes the exponent for leading finite-size corrections. We perform least-squares fits with $\kappa=1.16, 1.18, 1.2$ and $L=48, 64, 96, 128, 192$. We consider the situations with $y_t$ being free or fixed at $0.608$ and $0.58$, which were estimated for the special transition of classical O(2) model in spin~\cite{deng2005surface} and flow~\cite{unpublished} representations, respectively. For each situation, we obtain reasonably good results for large $L_{\rm min}$. When the leading correction term is present, the best estimate of $\omega_1$ is $\omega_1 \approx 1.4$ (Table~\ref{TABLE_S0}), which is larger than $\omega_1=0.789$ of 3D O(2) value~\cite{guida1998critical} and $\omega_1=1$ originating from boundary irrelevant fields~\cite{toldin2020boundary}, indicating that the correction term with $\omega_1 \le 1$ is either absent or weak. Hence, as shown in Table~\ref{TABLE_S}, we also perform fits without incorporating correction term, which have a reduced number of fitting parameters, and examine the stability of fitting results by varying $L_{\min}$. By comparing the fits, our final estimate of $\kappa_c$ is $\kappa_c=1.18(2)$.

\headline{Kosterlitz-Thouless-like critical phase}
We explore the critical phase on the large-$t$ side of Kosterlitz-Thouless-like transition for $\kappa=10$. For each $t$ in the set \{$0.027$, $0.03$, $0.035$, $0.04$, $0.045$, $0.05$\}, we perform scaling analyses for $g(L/2)$ according to Eq.~(\ref{eqfitBOK}) with $b=-\eta$, which corresponds to the leading FSS. The results are summarized in Table~\ref{TABLE_K}, which demonstrates that the fits are precise only at large sizes. Moreover, as $t$ increases, the exponent $\eta$ decreases.

\headline{Ordinary critical phase}
We analyze the ordinary critical phase at $\kappa=0.4$ and $t=t_c$. We fit $g(L/2)$ to Eq.~(\ref{eqfitBOK}) with $b=2y_h-4$. The results are presented in Table~\ref{TABLE_O}. For $L_{\rm min}=48$, $64$ and $96$, we find $b=-2.41(2)$, $-2.45(4)$ and $-2.5(1)$ with $\chi^2$/DF $\approx 1.3$, $1.1$ and $1.4$, respectively. These results are compatible with the exponent $2y_h-4$ with $y_h=0.781(2)$ of the classical O(2) ordinary surface criticality~\cite{deng2005surface}. If a correction term is included and the fitting ansatz becomes $g(L/2)=L^b(a+c L^{-\omega_1})$, the effects from corrections decrease rapidly with $L$ as $L^{b-\omega_1}$. It is practically difficult to estimate the amplitude of finite-size corrections.

\headline{Extraordinary critical phase}
We analyze the FSS for the extraordinary phase. We fit $g(L/2)$ to
\begin{equation}
	g(L/2) =a [{\rm ln}(L/l_0)]^{-\hat{q}}.
		\label{eqfitE1}
\end{equation}	
The results are given in Table~\ref{TABLE_E1}. If $\hat{q}$ is free, we obtain $0.3 \lessapprox \hat{q} \lessapprox 0.7$ and find that $l_0$ drastically decreases upon increasing $\kappa$. When $\hat{q}=0.59$ is fixed, we obtain stable fitting results of $a$ and $l_0$ for each considered $\kappa$. For $l_0$, instance results are $l_0=0.31(3)$, $0.21(1)$, $0.04(4)$, $0.0108(5)$ and $0.002(1)$ with $\chi^2$/DF $\approx$ $0.3$, $1.8$, $0.9$, $0.7$ and $0.5$, for $\kappa=2$, $3$, $5$, $7$ and $10$, respectively.

Assuming the existence of extraordinary-log critical universality, for each $\kappa$, we fit the data of $\rho_s$ to
\begin{equation}
\rho_s L =a + b {\rm ln}L.
\end{equation}
We obtain preferred fits with $L_{\rm max}=192$ for the deep extraordinary regime. For $\kappa=5$, we obtain $b=1.14(3)$ with $L_{\rm min}=64$ and $\chi^2/{\rm DF} \approx 0.9$. For $\kappa=7$, we obtain $b=1.15(3)$ with $L_{\rm min}=64$ and $\chi^2/{\rm DF} \approx 2.8$. For $\kappa=10$, we obtain $b=1.1(1)$ with $L_{\rm min}=96$ and $\chi^2/{\rm DF} \approx 0.7$. To obtain a unique estimate of fitting parameters, we analyze the sum of the scaled superfluid stiffness $\rho_s L$ over $\kappa=2$, $3$, $5$, $7$ and $10$ by performing fits to
\begin{equation}
\sum_{\kappa} \rho_s L =A + B {\rm ln}L.
\label{eqfitE2}
\end{equation}	
As summarized in Table~\ref{TABLE_E2}, we obtain reasonably good fits with $\chi^2/{\rm DF} \sim 1$ for $L_{\rm max}=192$ and $128$. For $L_{\rm max}=192$, we obtain $A=-5.3(7)$, $B=5.8(2)$ and $\chi^2/{\rm DF} \approx 2.0$ with $L_{\rm min}=64$, as well as $A=-8.8(2.0)$, $B=6.5(4)$ and $\chi^2/{\rm DF} \approx 0.5$ with $L_{\rm min}=96$. For $L_{\rm max}=128$, we obtain $A=-4.6(8)$, $B=5.6(2)$ and $\chi^2/{\rm DF} \approx 0.8$ with $L_{\rm min}=64$.

\begin{table*}
\begin{center}
\caption{Fits of $R_{[10]}$ to Eq.~(\ref{eqfitS}) for the special transition.}
\label{TABLE_S0}
\begin{tabular}{p{1.1cm}p{1.5cm}p{1.5cm}p{1.5cm}p{1.5cm}p{1.65cm}p{1.5cm}p{1cm}}
\hline
\hline
$L_{\rm min}$  & $\chi^2$/DF  & $\kappa_c$&$y_t$&$R_{[10]}^c$&$a_1$&$b_1$&$\omega_1$\\
\hline

48  &5.14/9&1.12(7)&0.50(5) &0.02(22)&0.06(1)&0.4(2)&0.4(7)\\
64  &4.47/6&1.1(1)&0.55(8) &0.1(3)&0.05(2)&0.7(5.2)&0.7(3.0)\\
96  &2.92/3&1.15(3)&0.4(1) &0.09(3)&0.09(6)&6.24(1)&1.4(2)\\
\hline

48  &8.83/10&1.13(5)&0.608 &0.03(18)&0.0390(9)&0.4(2)&0.4(7)\\
64  &5.03/7&1.1(1)&0.608 &0.1(4)&0.038(1)&0.4(2.3)&0.5(2.9)\\
96  &4.41/4&1.16(2)&0.608 &0.10(2)&0.038(1)&8.900(6)&1.5(2)\\

\hline

48  &7.09/10&1.13(5)&0.58 &0.03(19)&0.044(1)&0.4(2)&0.4(7)\\
64  &4.62/7&1.1(1)&0.58 &0.1(3)&0.043(1)&0.5(3.4)&0.6(3.0)\\
96  &3.98/4&1.16(2)&0.58 &0.10(2)&0.043(2)&7.903(7)&1.4(2)\\

\hline
\hline
\end{tabular}
\end{center}

\begin{center}
\caption{Fits of $R_{[10]}$ to Eq.~(\ref{eqfitS}) for the special transition with $b_1=0$.}
\label{TABLE_S}
\begin{tabular}{p{1.5cm}p{2cm}p{2cm}p{2cm}p{2cm}p{1.2cm}}
\hline
\hline
$L_{\rm min}$  & $\chi^2$/DF  & $\kappa_c$&$y_t$&$R_{[10]}^c$&$a_1$\\
\hline
48  &108.12/11&1.25(1)&0.29(5) &0.160(7)&0.15(3)\\
64  &37.02/8&1.206(7)&0.44(8) &0.138(4)&0.08(3)\\
96 &4.41/5&1.184(6)&0.4(1) &0.123(4)&0.10(7)\\
128  &0.33/2&1.175(5)&0.8(3) &0.117(4)&0.01(2)\\
\hline
48  &145.25/12&1.206(2)&0.608 &0.1398(8)&0.0394(9)\\
64  &41.39/9&1.197(2)&0.608 &0.133(1)&0.037(1)\\
96  &6.35/6&1.180(3)&0.608 &0.121(2)&0.038(1)\\
128  &0.84/3&1.175(7)&0.608 &0.116(5)&0.035(2)\\
\hline
48  &138.73/12&1.208(2)&0.58 &0.1407(8)&0.045(1)\\
64  &40.01/9&1.198(2)&0.58 &0.134(1)&0.042(1)\\
96  &5.84/6&1.181(4)&0.58 &0.121(2)&0.043(2)\\
128  &0.98/3&1.175(7)&0.58&0.116(5)&0.040(2)\\
\hline
\hline
\end{tabular}
\end{center}
\end{table*}

\begin{table*}
\begin{center}
\caption{Fits of $g(L/2)$ to Eq.~(\ref{eqfitBOK}) for the large-$t$ side of Kosterlitz-Thouless-like transition at $\kappa=10$.}
\label{TABLE_K}
\begin{tabular}{p{1.5cm}p{1.5cm}p{2cm}p{2cm}p{1.4cm}}
\hline
\hline	
$t$ &$L_{\rm min}$ &  $\chi^2$/DF&$a$ &$b$\\
\hline
0.027
&48  &208.63/3&1.115(3) &--0.1700(6)\\
&64  &36.28/2&1.080(4) &--0.1628(8)\\
&96  &0.96/1&1.01(1)&--0.150(2) \\
\hline
0.03
&48  &591.62/3&1.016(2)&--0.1264(4) \\
&64  &127.87/2&0.980(2) &--0.1187(5)\\
&96  &6.02/1&0.929(5) &--0.108(1)\\
\hline
0.035
&48  &437.57/3&0.990(1)&--0.0979(3) \\
&64  &119.14/2&0.964(2) &--0.0924(4)\\
&96  &5.54/1&0.925(4)&--0.0841(9) \\
\hline
0.04
&48  &96.45/3&1.010(2) &--0.0881(5)\\
&64  &28.96/2&0.991(3)&--0.0839(7) \\
&96  &0.65/1&0.950(8) &--0.075(2)\\	
\hline
0.045
&48  &24.06/3&1.017(3)&--0.0788(7) \\
&64  &6.70/2&1.003(4) &--0.076(1)\\
&96  &1.05/1&0.97(1)&--0.069(3) \\
\hline
0.05
&48  &18.18/3&1.017(4) &--0.070(1)\\
&64  &9.31/2&1.004(6) &--0.067(1)\\
&96  &0.98/1&0.96(2)&--0.058(4) \\
\hline
\hline					
\end{tabular}
\end{center}
\end{table*}

\begin{table*}
\begin{center}
\caption{Fits of $g(L/2)$ to Eq.~(\ref{eqfitBOK}) for the ordinary critical phase at $\kappa=0.4$.}
\label{TABLE_O}
\begin{tabular}{p{1.4cm}p{1.5cm}p{2cm}p{2cm}p{1.4cm}}
\hline
\hline
$L_{\rm max}$&$L_{\rm min}$ &  $\chi^2$/DF &$a$&$b$\\
\hline
192&32 &7.02/4&66.6(1.8)&--2.374(7) \\
&48  &4.01/3&76.5(6.5)&--2.41(2) \\
&64  &2.15/2&90.9(14.1)&--2.45(4) \\
&96  &1.41/1&141.2(77.8)&--2.5(1) \\
128&32 &2.62/3&66.2(1.8)&--2.373(7) \\
&48  &0.84/2&73.9(6.4)&--2.40(2) \\
&64  &0.002/1&83.7(13.7)&--2.43(4) \\
\hline
\hline
\end{tabular}
\end{center}

\begin{center}
\caption{Fits of $g(L/2)$ to Eq.~(\ref{eqfitE1}) for the extraordinary phase at $\kappa=2$, $3$, $5$, $7$ and $10$.}
\label{TABLE_E1}
\begin{tabular}{p{1cm}p{1.8cm}p{2.5cm}p{2.5cm}p{2.5cm}p{1cm}}
\hline
\hline
$\kappa$ &$L_{\rm min}$ &  $\chi^2$/DF &$a$&$l_0$&$\hat{q}$  \\
\hline
2&16  &1.93/4&0.68(1) &3.5(3)&0.32(1)\\
&32  &0.12/3&0.76(8) &2.2(9)&0.38(5)\\
&48  &0.07/2&0.7(2) &3.3(4.9)&0.3(2)\\
&64  &0.03/1&0.8(9) &1.7(8.1)&0.4(5)\\

&16  &177.76/5&1.170(2) &0.648(8)&0.59\\
&32  &8.09/4&1.248(7) &0.40(2)&0.59\\
&48  &0.89/3&1.29(2) &0.31(3)&0.59\\
&64  &0.12/2&1.31(4) &0.25(7)&0.59\\
&96  &0.07/1&1.33(8) &0.2(1)&0.59\\

\hline
3&16  &2.74/4&1.11(8) &1.0(2)&0.42(3)\\
&32  &1.01/3&0.9(1) &2.3(1.3)&0.33(6)\\
&48  &0.42/2&1.5(2.0) &0.3(1.2)&0.5(5)\\

&16  &16.72/5&1.649(3) &0.257(4)&0.59\\
&32  &7.21/4&1.69(1) &0.21(1)&0.59\\
&48  &0.42/3&1.73(2) &0.16(2)&0.59\\
&64  &0.31/2&1.75(5) &0.15(4)&0.59\\
&96  &0.003/1&1.7(1) &0.2(2)&0.59\\

\hline
5&16  &2.17/4&2.7(9) &0.03(3)&0.7(1)\\
&32  &1.30/3&1.5(6) &0.3(5)&0.5(2)\\
&48  &1.14/2&2.7(6.5) &0.02(22)&0.7(8)\\
&64  &0.99/1&1.1(1.3) &1.2(7.4)&0.3(5)\\

&16  &2.57/5&2.221(6) &0.053(1)&0.59\\
&32  &1.72/4&2.20(2) &0.058(5)&0.59\\
&48  &1.15/3&2.23(4) &0.05(1)&0.59\\
&64  &1.08/2&2.21(7) &0.05(2)&0.59\\
&96  &0.85/1&2.3(2) &0.04(4)&0.59\\

\hline
7&16  &2.29/4&4.4(2.7) &0.001(4)&0.7(2)\\
&32  &1.34/3&1.7(9) &0.1(3)&0.4(2)\\

&16  &3.31/5&2.692(9) &0.0108(5)&0.59\\
&32  &1.67/4&2.66(3) &0.013(2)&0.59\\
&48  &0.83/3&2.70(5) &0.010(3)&0.59\\
&64  &0.06/2&2.63(9) &0.015(7)&0.59\\
&96  &0.001/1&2.6(2) &0.02(2)&0.59\\

\hline
10&16  &12.53/5&3.35(2) &0.00084(8)&0.59\\
&32  &0.93/4&3.22(4) &0.0017(4)&0.59\\
&48  &0.91/3&3.21(8) &0.0018(7)&0.59\\
&64  &0.91/2&3.2(1) &0.002(1)&0.59\\
&96  &0.08/1&3.0(3) &0.01(1)&0.59\\						
\hline
\hline					
\end{tabular}
\end{center}
\end{table*}

\begin{table*}
\begin{center}
\caption{Fits of the summed scaled stiffness $\sum \rho_s L $ over $\kappa=2,3,5,7$ and $10$ to Eq.~(\ref{eqfitE2}) for the extraordinary phase.}
\label{TABLE_E2}
\begin{tabular}{p{1.5cm}p{1.6cm}p{2cm}p{2cm}p{1cm}}
\hline
\hline				
$L_{\rm max}$&$L_{\rm min}$ &  $\chi^2$/DF &$A$&$B$ \\
\hline
192&32  &95.84/4&0.2(2) &4.47(5)\\
&48  &28.71/3&--2.3(4) &5.10(9)\\
&64  &3.98/2&--5.3(7) &5.8(2)\\
&96  &0.45/1&--8.8(2.0) &6.5(4)\\
\hline
128&32  &64.32/3&0.5(2) &4.41(5)\\
&48  &16.33/2&--1.9(4) &5.0(1)\\
&64  &0.77/1&--4.6(8) &5.6(2)\\
\hline
\hline
\end{tabular}
\end{center}
\end{table*}

\end{document}